\documentclass[preprint,12pt]{elsarticle}

\usepackage{lineno,hyperref} 
\modulolinenumbers[1] 

\usepackage{amsmath,amssymb}

\usepackage{bm}

\usepackage{float}

\usepackage{longtable}
\usepackage{adjustbox}
\usepackage{latexsym}

\usepackage{textcomp,marvosym}

\usepackage{coolcaption}

\usepackage{color}
\newcommand{\blue}{\color[rgb]{0,0,1}}

\definecolor{pcbblue}{rgb}{0,0,0.6}
\definecolor{gray2}{rgb}{0.9,0.9,0.9}

 \newtheorem{Proposition}{Proposition}
 \newtheorem{Lemma}{Lemma}

\journal{Commun Nonlinear Sci and Numer Simulat}

\begin{document}
\begin{frontmatter}


\title{CAR T cell therapy in B-cell acute lymphoblastic leukaemia: Insights from mathematical models}







\author[a,* ]{Odelaisy Le\'on-Triana}
\author[b,*]{Soukaina Sabir}
\fntext[myfootnote]{These two authors contributed equally to this work.}
\author[a]{Gabriel F. Calvo}
\author[a]{Juan Belmonte-Beitia}
\author[c]{Salvador Chuli\'an}
\author[c]{\'Alvaro Mart\'{\i}nez-Rubio}
\author[c]{Mar\'{\i}a Rosa}
\author[d,e]{Antonio P\'erez-Mart\'{\i}nez}
\author[f]{Manuel Ramirez-Orellana}
\author[a]{V\'{\i}ctor M. P\'erez-Garc\'{\i}a}

\address[a]{Department of Mathematics, Mathematical Oncology Laboratory (MOLAB), Universidad de Castilla-La Mancha, Ciudad Real, Spain}
\address[b]{Faculty of Sciences, University Mohammed V, Rabat, Morocco}
\address[c]{Department of Mathematics, Universidad de C\'{a}diz, Biomedical Research and Innovation Institute of Cádiz (INiBICA), Hospital Universitario Puerta del Mar, C\'{a}diz, Spain}
\address[d]{Translational Research Unit in Paediatric Haemato-Oncology, Haematopoietic Stem Cell Transplantation and Cell Therapy, Hospital Universitario La Paz, Madrid, Spain}
\address[e]{Paediatric Haemato-Oncology Department, Hospital Universitario La Paz, Madrid, Spain}
\address[f]{Department of Paediatric Haematology and Oncology, Hospital Infantil Universitario Ni\~{n}o Jes\'us, Universidad Aut\'onoma de Madrid, Madrid, Spain}

\linenumbers 

\begin{abstract}
Immunotherapies use components of the patient immune system to selectively target cancer cells. The use of chimeric antigenic receptor (CAR) T cells to treat B-cell malignancies  --leukaemias and lymphomas-- is one of the most successful examples, with many patients experiencing long-lasting full responses to this therapy. This treatment works by extracting the patient's T cells and transducing them with the CAR,  enabling them to recognize and target cells carrying the antigen CD19$^+$, which is expressed in these haematological ancers. 
\par
Here we put forward a mathematical model describing the time response of leukaemias to the injection of CAR T cells. The model accounts for mature and progenitor B-cells, leukaemic cells, CAR T cells and side effects by  including the main biological processes involved. The model explains the early post-injection dynamics of the different compartments and the fact that the number of CAR T cells injected does not critically affect the treatment outcome. An explicit formula is found that gives the maximum CAR T cell expansion {\em in vivo} and the severity of side effects. Our mathematical model captures other known features of the response to this immunotherapy. It also predicts that CD19$^+$ cancer relapses could be the result of competition between  leukaemic  and CAR T cells, analogous to predator-prey dynamics. We discuss this in the light of the available evidence and the possibility of controlling relapses by early re-challenging of the leukaemia cells with stored CAR T cells.
\par
\end{abstract}


\end{frontmatter}


\section{Introduction}
\label{Sec:Intro}

Cancer immunotherapy approaches use components of a patient's own immune system to selectively target cancer cells.  Immunotherapies are already an effective treatment option for several cancers due to their selectivity, long-lasting effects, and benefits for overall survival~{\blue\cite{Koury2018}}. 
\par

Chimeric antigen receptor (CAR) {T cell} therapy represents a major step in personalised cancer treatment, and is the most successful type of immunotherapy. 
This therapy is predicated on  the use of gene-transfer technology to instruct T lymphocytes to recognise and kill cancer cells. CARs are synthetic receptors that mediate antigen recognition, T cell activation, and { co-stimulation} to { increase} T cell functionality and persistence.  For { the} clinical application, { the} patient's T cells are obtained, genetically engineered { \em ex vivo} to express the synthetic receptor,  expanded and infused back into the patient~{\blue\cite{Sadelain}}.
\par

Clinical trials have shown promising results in end-stage patients with B-cell malignancies due to their expression of the CD19 protein~{\blue\cite{Cell}}. CAR T cells engineered to { recognise} this antigen have led to an early clinical response of up to 92\% in Acute Lymphoblastic Leukaemia (ALL) patients~{\blue\cite{3,ALL1,Pan2017,Militou}}. Good results have been reported for large B-cell lymphomas~{\blue\cite{Lymphoma1,Lymphoma2}} and multiple myelomas~{\blue\cite{MMyeloma}}. These successes have led to the approval of CAR T therapies { for use} against CD19 for treatment of B-ALL and diffuse large B-cell lymphomas~{\blue\cite{Cell}}. 
\par

CAR T-related toxicities are {cytokine release syndrome} (CRS), due to the release of cytokines during CAR T cell action, and immune effector-cell-associated neurotoxicity syndrome (ICANS). CRS symptoms including hypotension, pulmonary { oedema}, multiorgan failure, and even death, are {now better controlled using IL-6 inhibitor tocilizumab}~{\blue\cite{CRS}}.
\par

{Despite the success of { CAR T cell} therapy { seen} so far, a variable fraction, between 30\% and 60\%, of  patients relapse after treatment. There are two different types of post-CAR { relapse}. In the first type, post-CAR leukaemic cells show expression of the CD19$^+$ antigen and other immunophenotypic characteristics that are the same as those of the original clone.
This is consistent with the recurrence of the initial leukaemic clone. In { this} case, pre-CAR and post-CAR blasts typically show the same CD19 expression levels.}

{{ This type of recurrence shows a down-regulation} of the CD19 antigen~{\blue\cite{Relapse1}}. 
In this situation, { CAR T cells} cannot { recognise} their targets and the { tumour} regrows. In contrast to CD19$^+$ { recurrence}, CD19$^-$ { recurrence occurs} despite functional persistence of { CAR T cells} and ongoing B-cell aplasia~{\blue\cite{Relapse1,NatureMedicine2019}}.}

\par

{ There are many previous studies} devoted to the mathematical modelling of { tumour-immune} cell interactions, see for instance~{\blue\cite{Eftimie2011,Starkov2014,Eftimie2016,Lopez2017,Konstorum2017,Mahlbacher2019}} and references therein. Mathematical models have the potential to provide a mechanistic understanding of { oncological} treatments, and may help in finding { optimised} strategies { to improve} treatment outcome~{\blue\cite{rev3,rev4}}.  This is why CAR T cell treatments have attracted the interest of mathematicians in the context of gliomas~{\blue\cite{Hedge,Sahoo}}, melanomas~{\blue\cite{Baar}} and B-cell malignancies~{\blue\cite{Kimmel,Rodrigues,Rodrigues2,Anna,Stein}}.
\par

In this { study}, we will describe mathematically the longitudinal dynamics of  B cells, leukaemic clones and CAR T cells. The mathematical models will be shown to provide both a mechanistic explanation { for} the results of different clinical trials and  formulas quantifying some of the observed phenomena. We will discuss some implications for CD19$^+$ relapses and how it { might be possible to control them} by { re-challenging} the  { cancer early}  with CAR T cells.
\par

\section{Mathematical models and parameter estimation}
\label{Sec:MathModels}

\subsection{Basic mathematical model}

Our mathematical model accounts for the evolution { over time} of several interacting cellular populations distributed into five compartments. Let $C(t)$, ${ L}(t)$, $B(t)$, $P(t)$, and $I(t)$ denote the { non-negative} time-varying functions representing the number of CAR T cells, { leukaemic} cells, mature healthy B cells, CD19$^-$ { haematopoietic} stem cells (HSCs), and CD19$^{+}$ B cell progenitors (i.e. Pre-B, Pro-B and immature bone marrow B cells), respectively. Our { initial} autonomous system of differential equations { is as follows:}
 \begin{subequations}
 \label{model1}
 \begin{eqnarray} \label{model11}
 \frac{dC}{dt} & = & \rho_C \left( { L} + B\right) C + \rho_{\beta} I C - \frac{1}{\tau_C} C, \\ \label{model12}
 \frac{d{ L}}{dt} & = & \rho_{ L} { L} - \alpha { L} C, \\ \label{model13}
 \frac{dB}{dt} & = & \frac{1}{\tau_I} I - \alpha  B C -  \frac{1}{\tau_B} B, \\ \label{model14}
 \frac{dP}{dt} & = & \rho_P \left( 2 a_Ps(t) - 1\right) P - \frac{1}{\tau_P} P, \\ \label{model15}
 \frac{dI}{dt} & = & \rho_I \left( 2 a_Is(t) - 1\right) I  - \frac{1}{\tau_I}I + \frac{1}{\tau_P} P  - \alpha \beta I C. 
 \end{eqnarray}
Equation (\ref{model11}) involves two proliferation terms of CAR T cells due to stimulation by encounters with their target cells: either ${ L}(t)$, $B(t)$ or  $I(t)$. The parameter $\rho_C>0$ measures the { stimulation to mitosis} after encounters with CD19$^+$ .{ cells disseminated throughout the whole body} (mostly in the circulatory system). The parameter $\rho_{\beta}  = \beta \rho_C$, where $0<\beta<1$, accounts for the fact that immature B cells are located mostly in the bone marrow and encounters with CAR T cells will be less frequent. The last term describes the decay of CAR T cells with a mean lifetime $\tau_C$.
\par 
 
In contrast { to previous modelling} approaches~{\blue\cite{Kimmel,Rodrigues,Anna}}, we exclude a death term in the CAR T cell compartment due to interaction with target cells. { This is because the CAR T cells do not undergo apoptosis after killing the target cell~{\blue\cite{SerialKillers1,SerialKillers2}}. Also, { unlike in those} models, { there is no} {\em standard} proliferation term proportional to the population of CAR T cells, since these cells do not divide spontaneously~{\blue\cite{Tough1995}}; instead their clonal expansion is directly dependent { on stimulation} with the CD19 antigen. 
\par

{ Leukaemic} cells [see Eq. (\ref{model12})] have a net proliferation rate $\rho_{ L}>0$ and die due to encounters with CAR T cells. The parameter $\alpha$ measures the probability (per unit time and cell) { of an encounter between CAR T and CD19$^{+}$ cells}. { $\alpha$ and $\rho_C$ will, in general, be different,} due to possible asymmetric cell interactions. Namely, if $\alpha > \rho_C$, this would imply that CAR T cells kill CD19$^{+}$ target cells relatively faster than their own proliferation rate per target cell encountered. In contrast, if $\alpha < \rho_C$ then, on average, the killing process would be slower than the proliferation rate per target cell encountered by CAR T cells. For completeness, we will consider both cases via the { dimensionless} parameter $k = \rho_C/\alpha$. Other processes, encompassed by the term $- \alpha { L} C$, would include target cell recognition, killing and detachment, which are relatively much faster than the complete {\em rendezvous} kinetics~{\blue\cite{KillingCAR}}. { Our model implicitly  assumes that all T lymphocytes in the CAR product have a similar cell killing capacity. This could be the case if both CD4$^+$ T helper cells and CD8$^+$ T cells had a similar cell killing capacity, or if most of the CAR product contains CD8$^+$ cells.} 
{ A more complex mathematical framework should incorporate potential differences in killing capacity of these two T lymphocytes}~{\blue\cite{KillingCAR,Liadi2015}}.}
\par 
 
Equations (\ref{model13})-(\ref{model15}), { which} involve B cells, consist of a compartment for CD19$^{-}$ HSCs  (i.e. $P(t)$) { with} an asymmetric division rate $a_P$ and a differentiation rate $1/\tau_P$ into a new compartment accounting for all of the other CD19$^{+}$ differentiated states of bone marrow progenitor B cells (Pro-B, Pre-B, and immature cells) (embodied in $I(t)$). These cells { are} the source of mature B cells. Since all cells in the $I(t)$ compartment already express the CD19$^{+}$ antigen, they will be targets for the fraction of CAR T cells in the bone marrow, namely $\beta C(t)$. Finally, mature B cells $B(t)$, which cannot subsequently proliferate, are the terminal differentiation stage of these cells. They have a mean lifetime $\tau_B$, which is present in the last term of Eq. (\ref{model13}). The structure of the two { haematopoietic} compartments is similar to that proposed in { previous studies with} { haematopoiesis} models~{\blue\cite{Marciniak}}. In line with those models, the { signalling} function $s(t)$ can be assumed  to be of the saturable form $s(t) = 1/\left[1+k_s \left(P + I \right)\right]$, with $k_s >0$.
\par
   \begin{figure}
	\centering
	\includegraphics[width=0.9\columnwidth]{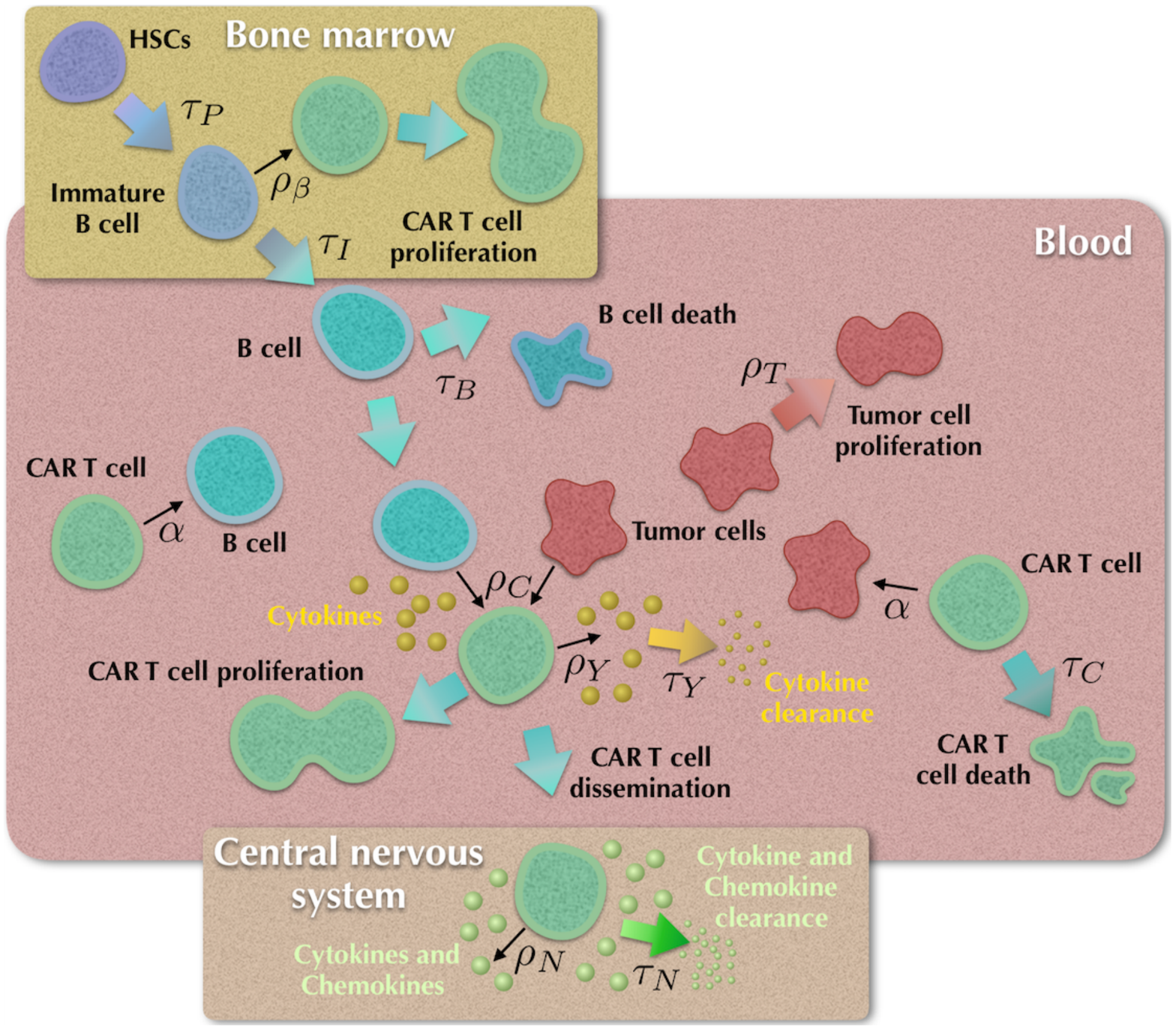}
	\caption{\textbf{Processes included in the mathematical model (\ref{model1})}. Mature B lymphocytes are generated from the CD19$^-$ { haematopoietic} stem cells (HSCs) and through differentiation of immature CD19$^+$ progenitors with characteristic lifetimes $\tau_P$ and $\tau_I$, respectively. CAR T cells are stimulated when meeting CD19$^+$ B cells (normal, leukaemic or immature) with stimulation parameters $\rho_C$ and $\rho_\beta$, and undergo apoptosis with a lifetime $\tau_C$. { leukaemic} cells proliferate with a rate $\rho_{ L}$. Both mature B and { leukaemic} cells are destroyed via encounters with the CAR T cells with a killing efficiency $\alpha$. Cytokines are released with a rate $\rho_Y$, { which} may result in acute toxicities (cytokine release syndrome) and are cleared at a rate $1/\tau_Y$. A fraction of the CAR T cells disseminates and infiltrates the central nervous system. Typical neurological toxicities induced by CAR T cells are immune effector cell-associated neurotoxicity {  syndromes}. { Thick coloured arrows describe different individual cell processes. Narrow coloured arrows depict both release and degradation of chemical agents. Black double-headed arrows indicate cell-cell interactions.}}
\label{themodel}
\end{figure}

To describe CRS, let us define a variable $Y(t)$ { for} the cytokines released upon stimulation of CAR T cells by the antigens with rate $\rho_Y$ and { cleared with} rate $1/\tau_Y$. ICANS will be related to the number of { CAR T cells} infiltrating the central nervous system (CNS), and is expected to be proportional to the total number of { CAR T cells}, with a proportionality coefficient $\rho_N$, and removed at a rate $1/\tau_N$. Thus, toxicity can be { described by} the following equations:
\begin{eqnarray} \label{model16}
\frac{dY}{dt} & = & \rho_Y C \left( { L} + B\right) - \frac{1}{\tau_Y} Y, \\ \label{model17}
 \frac{dN}{dt} & = & \rho_N C - \frac{1}{\tau_N} N. 
 \end{eqnarray}
 \end{subequations}
Notice that Eqs. (\ref{model11})-(\ref{model15}) are uncoupled from Eqs. (\ref{model16})-(\ref{model17}). A schematic summary of the biological processes encompassed by our basic mathematical model (\ref{model11}-\ref{model17}) is shown in Fig. \ref{themodel}.
\par

\ref{AppendixA} shows some mathematical results on { the} existence, uniqueness and positiveness of the solutions of system \eqref{model11}-\eqref{model15}.

\subsection{Reduced mathematical models}

Equations (\ref{model11})-(\ref{model15}) exclude different biological facts such as heterogeneity in the CAR T-lymphocyte subpopulations, the differential expression of the CD19 antigen { over} { leukaemic} and healthy B cells subclones, the role of regulatory T-cells, etc. However, { there are still many} parameters to be determined. The contribution of the bone marrow Eqs. (\ref{model14})-(\ref{model15}) is to account for the generation of new B-cells.  Hence, to capture their role while simplifying the full system, we can compute the equilibria for Eqs. (\ref{model14})-(\ref{model15})
\par
\begin{equation}\label{equi}
I = \frac{\frac{1}{\tau_Pk_S} \left(\frac{2a_P\tau_P\rho_P}{1+\tau_P\rho_P}-1\right)}{\frac{1}{\tau_I} + \frac{1}{\tau_P} + \rho_I\!\left[1-\frac{a_I}{a_P}\left(1+\frac{1}{\tau_P \rho_P}\right)\right] + \alpha \beta C} \equiv \frac{I_0}{1+C/C_{50}} \, ,
\end{equation}
and assume (\ref{equi}) to hold for all { time}. This provides a suitable representation of the contribution of immature B cells in the bone marrow { to global} disease dynamics. Then, Eqs. (\ref{model11})-(\ref{model15}) reduce to the set
 \begin{subequations}
 \label{model2}
 \begin{eqnarray} \label{model21}
 \frac{dC}{dt} & = & \rho_C \left( { L} + B \right) C + \frac{\rho_C \beta I_0}{1 + C/C_{50}} C - \frac{1}{\tau_C} C, \\ \label{model22}
 \frac{d{ L}}{dt} & = & \rho_{ L} { L} - \alpha { L} C, \\ \label{model23}
 \frac{dB}{dt} & = & \frac{I_0/\tau_I}{1+C/C_{50}} - \alpha  B C -  \frac{1}{\tau_B} B.
 \end{eqnarray}
 \end{subequations}
 \par
 
In the first weeks after CAR T injection, the main contribution to the dynamics is the expansion of these cells and their effect on the healthy B  and { leukaemic} cells. Thus, we may neglect the contribution of the { haematopoietic} compartments in Eqs. (\ref{model21}) and (\ref{model23}) to get
 \begin{subequations}
 \label{model3}
 \begin{eqnarray} \label{model31}
 \frac{dC}{dt} & = & \rho_C \left( { L} + B\right) C - \frac{1}{\tau_C} C, \\ \label{model32}
 \frac{d{ L}}{dt} & = & \rho_{ L} { L} - \alpha { L} C, \\ \label{model33}
 \frac{dB}{dt} & = & - \alpha  B C - \frac{1}{\tau_B} B.
 \end{eqnarray}
 \end{subequations}

The study of existence and uniqueness of solutions, together with the stability of the critical points for both systems, are presented in  \ref{AppendixB} and \ref{AppendixC}.

\subsection{Parameter estimation}

B-cell lymphocyte lifetime $\tau_{B}$ is known to be about 5-6 weeks~{\blue\cite{Fulcher97}}. These cells account for a variable fraction between 5\% and 20\%~{\blue\cite{Enciclo}} of the total lymphocyte number~{\blue\cite{Alberts}} leading to $>10^{\text{11}}$ B-lymphocytes in humans. { In this paper, since CAR T cells are injected after lymphodepleting treatment, { in most simulations we set} the initial number of B-lymphocytes to be 2.5 $\times 10^{10}$ { to account} for the effect of this treatment.}
\par

ALLs are fast-growing { cancers} with proliferation rates $\rho_{{ L}}$ of the order of several weeks~{\blue\cite{Marciniak,Skipper1970}}. Na\"ive CD8$^+$ T cells are quiescent, their mean lifetime ranges from months to years, and { they} enter the cell cycle following interaction with their antigen~{\blue\cite{Nayar2015,Kasakowski2018}}. These activated CD8$^+$ T cells induce cytolysis of the target cells and secrete cytokines such as TNF-$\alpha$ and IFN$\gamma$. Following activation, most effector cells undergo apoptosis after two weeks, with a small proportion of cells surviving to become CD8$^+$ memory T cells capable of longer { survival}~{\blue\cite{Nayar2015}}. Recent { studies} have reported longer survival values of about one month~~{\blue\cite{NatureMedicine2019}}. Thus, we will take the mean lifetime $\tau_{C}$ of CAR T cells { to be} in the range of 2-4 weeks.
\par
To estimate the interaction parameter $\alpha$ we will use the fact that when measured by flow cytometry or qPCR, CAR T cells in children treated for ALL reached a maximum {\em in vivo} expansion { at} around 14 days~{\blue\cite{Lee}}, { which} is a typical value observed in other clinical studies. Finally, the mitotic rate $\rho_C$, related to the stimulatory effect of each encounter { between T cells and the} CD19$^+$ cells, will be taken to be proportional to $\alpha$ { ($\rho_C=k \alpha$)}, { with $k\in(0.05,2)$. The exact value would depend on the properties of the { CAR T product,} but taking $B+L$ initially to be around $10^{11}$ and using Eqs. (\ref{model31}) we obtain an initial exponential growth rate for { CAR T cells of around} $k$ day$^{-1}${ , in line with values reported} in other models (e.g in Ref. \cite{Stein}, the authors obtained 0.89 day$^{-1}$ from data).}

{ The parameter values used in this paper are summarised} in Table \ref{table1}.
\par
	
\begin{table}[H]
	\begin{center}
		\begin{tabular}{|c|l|c|c|c|}
			\hline
			Parameter &   Meaning                            & Value                               & Units                     & Source 	             \\ \hline
			$\tau_B$  & B-lymphocyte          &  $30-60$                    		        & day \hbox{ }\hbox{ } &  \cite{Fulcher97} \\ 
					& lifetime			&							&				& \\ \hline        
			$\rho_{ L}$  & { Leukaemic} growth rates             &  $1/30-1/60$                		        & day$^{\text{-1}}$  &           \\ \hline
			$\tau_C$  & Activated CAR T           & $14 - 30$ 			& day \hbox{ }\hbox{ } & \cite{Nayar2015,NatureMedicine2019} \\ 
			                 & cell lifetime				     &						&				&			    \\ \hline
			 $\rho_C$ & Mitotic stimulation   & $ (0.05-2)   \times  \alpha$	& 	day$^{\text{-1}}$	& 			     \\
			 		& of CAR T cells by       &                                            &      $\times$ cell$^{\text{-1}}$                       &                            \\ 
					& CD19$^+$ cells           &                                      &                                   &                       \\ \hline 
			 $\alpha$ & Killing efficiency  &      $\sim$ 10$^{-\text{11}}$ 	& 	day$^{\text{-1}}$			& Estimated 			     \\
			 		& of CAR T cells 				     &                                           &         $\times$ cell$^{\text{-1}}$                      &    from \cite{Lee}                         \\  \hline 
			
				{  $k$}  & 	{   $\rho_C$ and $\alpha$ ratio}&  {      $ 0.05-2 $ }	& 		{  dimensionless }			&	{ Estimated and}		     \\
					& 		     &                                           &       &    { compatible with }                  \\ 
					&                                                                                         &                                &                      & { \cite{Stein}}  \\ \hline 

					{	$\tau_I$ } & { Immature bone marrow }            &   { $2-6 $ }               		        & { day } &   { 	\cite{Rolink, Shahaf}}               	        \\ 
					&  { B cell lifetime	}		&							&				& \\ \hline        
						{	$\beta$ } & { Fraction of CAR T cells   }            &   { $0.01-0.5 $ }               		        & 	{  dimensionless }&        { \cite{Yasuyuki} }               	   \\ 
					&  { in the bone marrow	}		&							&				& \\   \hline        
		\end{tabular}
		\caption{{ Relevant parameter values for} model Eqs. (\ref{model3})}
		\label{table1}
	\end{center}
\end{table}

\section{Results}
\label{Sec:Results}

In this section, we present the results obtained from systems \eqref{model2} and \eqref{model3}. { We first analyse} \eqref{model3}:

\subsection{Mathematical model (\ref{model3}) describes post CAR T cell injection dynamics} 

\begin{figure}[t!]
	\centering
	\includegraphics[width=0.9\columnwidth]{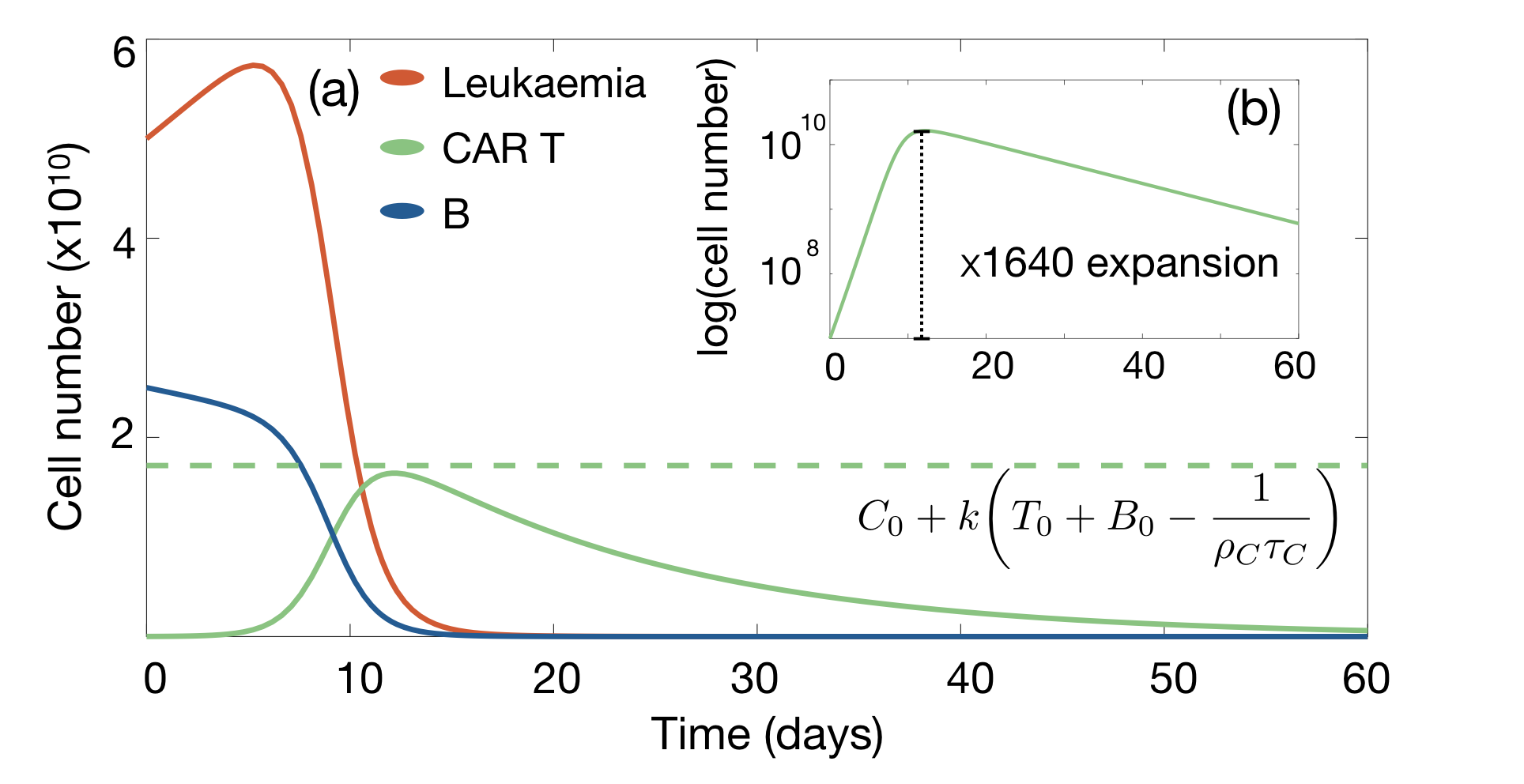}
	\vspace*{-5mm}
	\caption{\textbf{Typical dynamics of { leukaemic cell} (red curve), B-cell (blue curve) and  CAR T cell (green curve) compartments according to Eqs. (\ref{model3})}. (a) Simulations for parameters $\alpha$ = 4.5$\times$ 10$^{-\text{11}}$ day$^{-1}$cell$^{-1}$, 
$\tau_C$ = 14 days, $\rho_{ L}$ = 1/30 day$^{-\text{1}}$, $\rho_C = 0.25 \alpha$, $\tau_B$ = 60 days and injected cells $C_0$ = 10$^\text{7}$ corresponding, to 5$\times$10$^\text{5}$ cells per kg for a 20 kg child. Also, ${ L}_0 = 5 \times 10^{\text{10}}$ and $B_0 = 2.5 \times 10^\text{10}$, which correspond to typical values after lymphodepleting chemotherapy. (b) Logarithmic plot of the CAR T cell population.}
\label{prima}
\end{figure}

{ We first studied the dynamics of the system post-CAR T cell injection numerically,} as described by Eqs.~(\ref{model3}). Figure \ref{prima} shows a typical example. During the first two months { of the simulation,} CAR T cells { expanded, showing} a peak at about two weeks post-injection, before their numbers { stabilised and began to decrease.} Both the { leukaemic} and B-cell compartments experienced a continuous decrease towards undetectable values  representing the dynamics of a  patient without residual disease. The expansion of the CAR T population was exponential, increasing by several orders of magnitude [see Figure \ref{prima}(b)], in line with reported clinical experience and patient datasets~{\blue\cite{NatureMedicine2019}}.
\par

\subsection{The number of injected CAR T cells does not affect treatment outcome, but the stimulation rate does} 
\label{nCART}

\begin{figure}[t!]
	\centering
	\includegraphics[width=0.8\columnwidth]{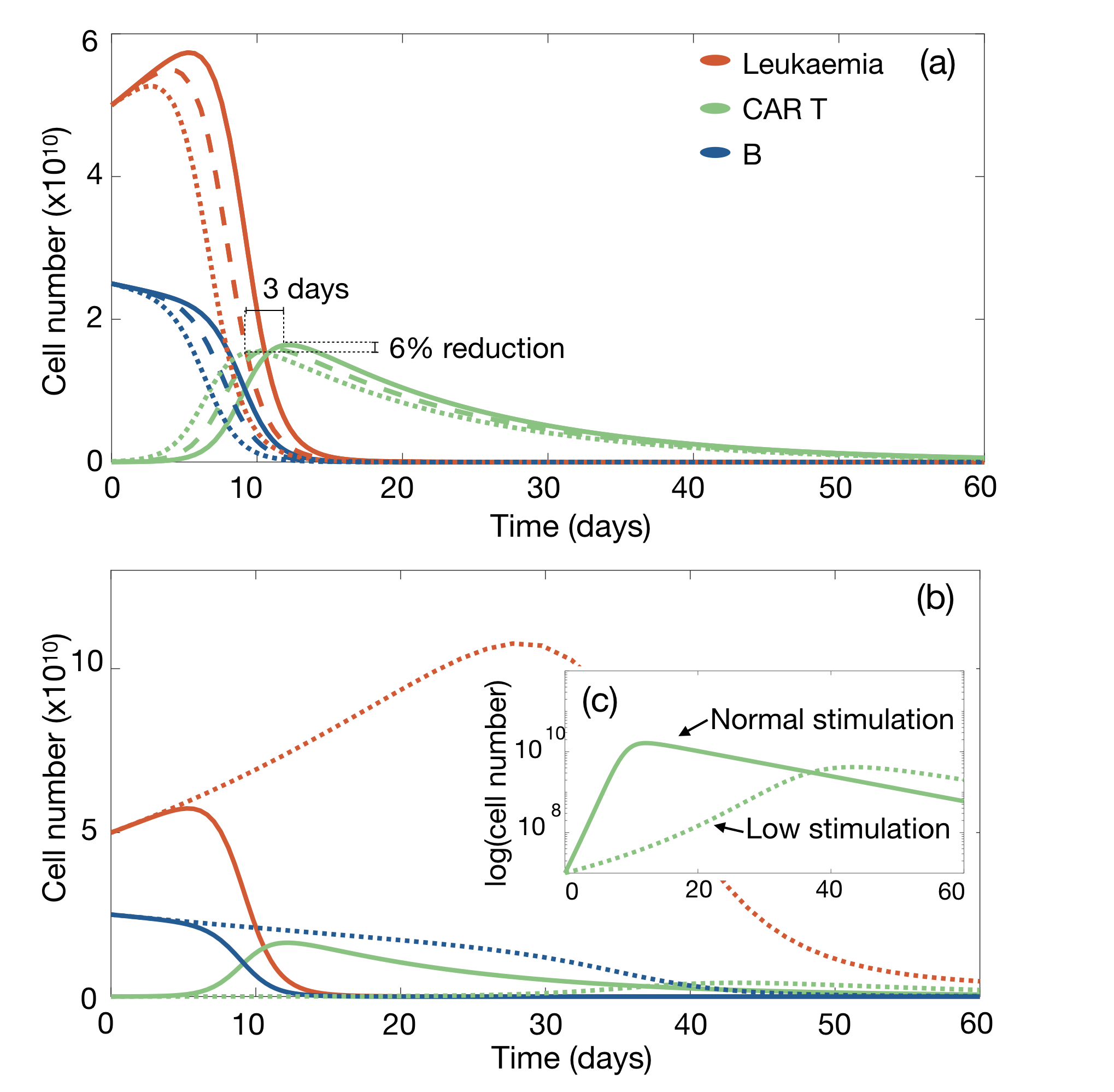}
	\vspace*{-3mm}
	\caption{\textbf{The number of injected CAR T cells does not affect treatment outcome, but the stimulation rate does.} (a) Dynamics of { leukaemic cells} (red line), B-cells (blue line) and CAR T cells (green line) according to Eqs. (\ref{model3}) for the virtual patient of Fig. \ref{prima} subject to injections of $5 \times 10^\text{5}$ cells/kg (solid lines), 
	$15 \times 10^\text{5}$ cells/kg (dashed lines) and $45 \times 10^\text{5}$ cells/kg (dotted lines). (b,c) Dynamics for stimulation rates $\rho_C = 0.25 \alpha$ (solid line) and $\rho_C = 0.05\alpha$ (dotted lines). (c)  { CAR T cell} expansion in log scale.}
\label{segunda}
\end{figure}

We next studied the dynamics of Eqs.~(\ref{model3}) under different numbers of injected CAR T cells. A typical example is displayed in Figure~\ref{segunda}(a). The change of one order of magnitude in the initial CAR T cell load resulted in minor changes in the maximum expansion achieved (of around $6\%$). A reduction in the time to peak expansion {\em in silico} of about $3$ days was observed. However, the persistence of CAR T cells was not affected by their initial load.
\par

{ It has already been observed} that when the number of CAR T cells seeded is small, the therapy can fail~{\blue\cite{Hartmann2017}}. We simulated {\em in silico}, { and saw that} the effect of a reduction in the growth efficiency of the cells (the stimulation rate $\rho_C$) and the dynamics { were} substantially affected [see Fig.~\ref{segunda}]. A { reduction in the efficiency of stimulation}  in the CAR T cells led to a slower growth of this population  {\em in silico}, resulting in { leukaemic cells} reaching { higher numbers} for almost two months without any clinical response.
\par

\subsection{Maximum expansion of CAR T cells { in vivo} and CRS}

System \eqref{model3} is amenable { to} finding useful analytical and semi-analytical expressions, which are all derived in \ref{AppendixD}. 
\par
At time $t_\textrm{max}$, which typically occurs within 2-4 weeks after injection of the CAR T cells, a first maximum in their number, denoted by $C_\textrm{max} \equiv C(t_\textrm{max})$, is achieved during the expansion phase. The value of $t_\textrm{max}$ can be calculated from the implicit relation
\begin{eqnarray} 
 \log\!\left[ \rho_{C}\tau_{C}\!\left( { L}_{0}\, e^{\rho_{ L} t_\textrm{max}} +  B_{0}\, e^{-\frac{t_\textrm{max}}{\tau_{B}}} \right)\right] - \alpha\int_{0}^{t_\textrm{max}}C(t)dt = 0 .
 \label{model3tmax} 
 \end{eqnarray}

Furthermore, it is possible to estimate the maximum number of CAR T cells $C_\textrm{max}$, which is approximately given by 
\begin{eqnarray} 
C_\textrm{max} \simeq C_{0} + k\!\left( { L}_{0} + B_{0} - \frac{1}{\rho_{C}\tau_{C}}\right) \! ,
 \label{model3CmaxApprox}
 \end{eqnarray}
where $C_{0}$, ${ L}_{0}$ and $B_{0}$ are the initial conditions for the CAR T, { leukaemic} and B cells (assumed to be positive), respectively. Thus, the maximum number of CAR T cells that can be { reached} during the first expansion phase will be related to the initial populations ${ L}_{0}$ and $B_{0}$ multiplied by the ratio $k$. { Note} also that, in practice, the contribution of $C_{0}$ is much smaller than the second term in Eq. (\ref{model3CmaxApprox}) and can be ignored, { suggesting} that the initial number of injected CAR T cells does not affect the peak, although it does contribute in (\ref{model3tmax}) when computing $t_\textrm{max}$. Both results (\ref{model3tmax}) and  (\ref{model3CmaxApprox}) provide an analytical justification of the numerical results discussed for Eqs.~(\ref{model3}) in Section \ref{nCART}.
\par

{ Explicit formulas for computing the { leukaemic} and B cell loads in patients at time $t_\textrm{max}$ can easily be obtained. These} are given by
 \begin{subequations} \label{model3TBmax}
 \begin{eqnarray} 
 { L}(t_\textrm{max}) &=& \frac{{ L}_{0}\, e^{\left( \rho_{ L} + \frac{1}{\tau_{B}}\right)t_\textrm{max}}}{\rho_{C}\tau_{C}\!\left(B_{0} + { L}_{0}\, e^{\left( \rho_{ L} + \frac{1}{\tau_{B}}\right)t_\textrm{max}}\right)}\, ,
 \label{model3Tmaxa} \\
 B(t_\textrm{max}) &=& \frac{B_{0}}{\rho_{C}\tau_{C}\!\left(B_{0} + { L}_{0}\, e^{\left( \rho_{ L} + \frac{1}{\tau_{B}}\right)t_\textrm{max}}\right)}\, .
 \label{model3Tmaxb} 
 \end{eqnarray}
 \end{subequations}
 Alternatively, if ${ L}(t_\textrm{max})$ and $B(t_\textrm{max})$ are available, together with ${ L}_{0}$, $B_{0}$, $\tau_{C}$ and $\tau_{B}$, then parameters $\rho_{C}$ and $\rho_{{ L}}$ can be estimated, { which is} relevant from a clinical viewpoint. 
 \par
 
\begin{figure}[t!]
	\centering
	\includegraphics[width=1\columnwidth]{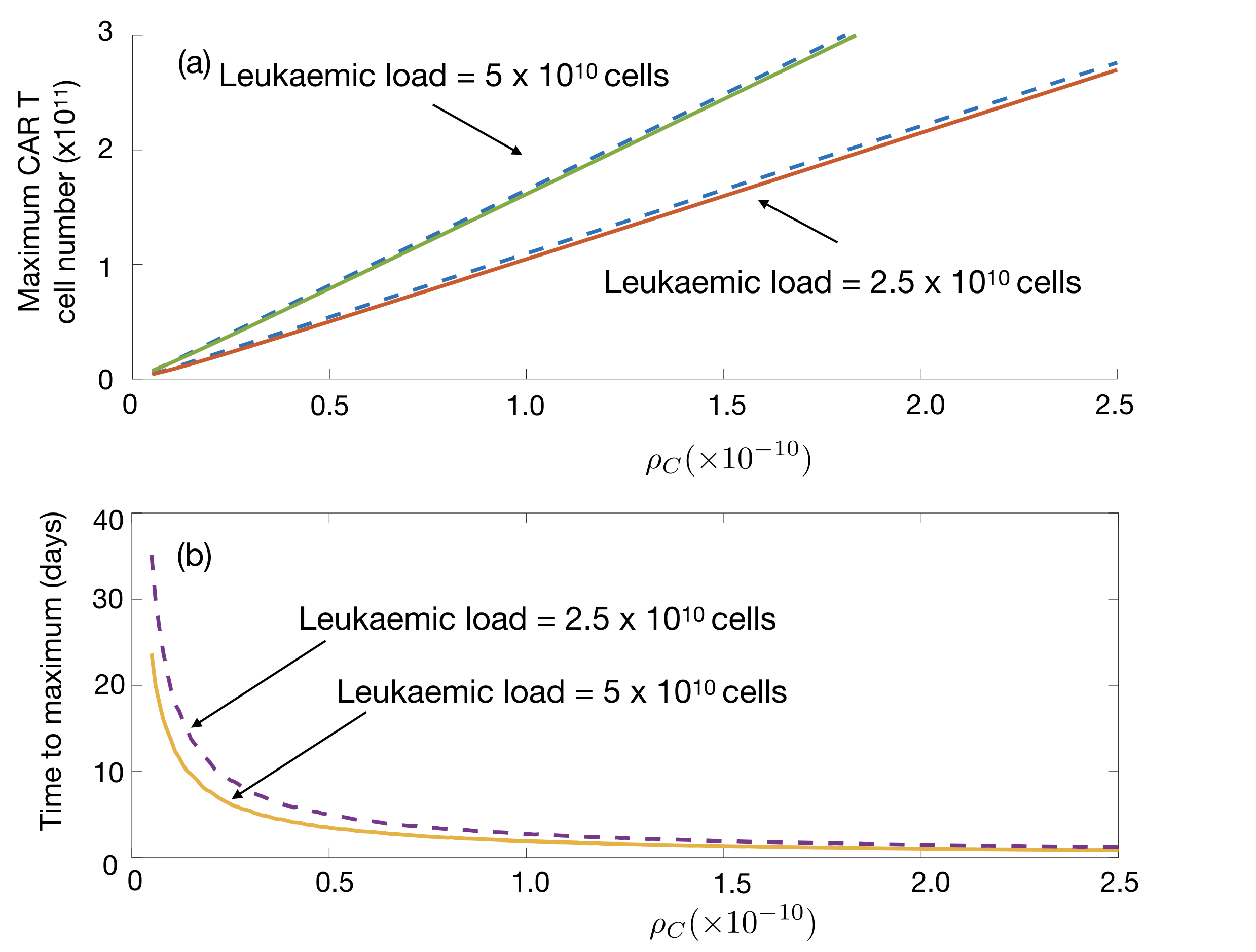}
	\vspace*{-3mm}
\caption{\textbf{Dependence on $\rho_C$ of the maximum number of CAR T cells and the time $t_\textrm{max}$ taken to achieve the maximum.} Common parameters for all plots are as in Figure \ref{prima}: $\alpha$ = 4.5$\times$ 10$^{-\text{11}}$ cell$^{-1}$ day$^{-1}$, 
	$\tau_C$ = 14 days, $\rho_{ L}$ = 1/30 day$^{-\text{1}}$, $\tau_B$ = 60 days and initial cell numbers $C_0$ = 10$^7$ cells, $B_0$= 2.5 $\times$ 10$^{\text{10}}$ cells. (a) Maximum value number of CAR T cells obtained for initial { leukaemic} loads of 5 $\times$ 10$^\text{10}$ cells (red) and  2.5 $\times$ 10$^\text{10}$ cells as a function of $\rho_C$. Solid line indicates the results obtained from Eqs. \ref{model3} and the dashed line the upper bound given by Eq. (\ref{model3CmaxApprox}). (b) Time to maximum value of CAR T cells for different { leukaemic} loads computed from Eq. (\ref{model3tmax}). }
\label{cmaxplot}
\end{figure}

Since toxicity, accounted for by Eqs.~(\ref{model16}) and (\ref{model17}), depends on the maximum CAR T cell number, one would expect a smaller ratio $k$ to lead to lower toxicities. Figure \ref{cmaxplot} shows the linear dependence of the maximum number of CAR T cells on $\rho_C$ as obtained from simulations of Eqs. (\ref{model3}), { which} is well approximated by Eq. (\ref{model3CmaxApprox}). The above result { points} to a { proportional relation} between the total { leukaemic} load and the severity of the CRS syndrome.
In fact, a strong correlation between the severity of CRS and disease { load} at the time of { CAR T cell} infusion has been noted in multiple clinical trials of { CAR T cell} therapy of { haematological} malignancies~{\blue\cite{Lee,Maude2,Davila,Turtle}}.

\subsection{CAR T cell persistence depends on the T cell mean lifetime} 

\begin{figure}[t!]
	\centering
	\includegraphics[width=0.9\columnwidth]{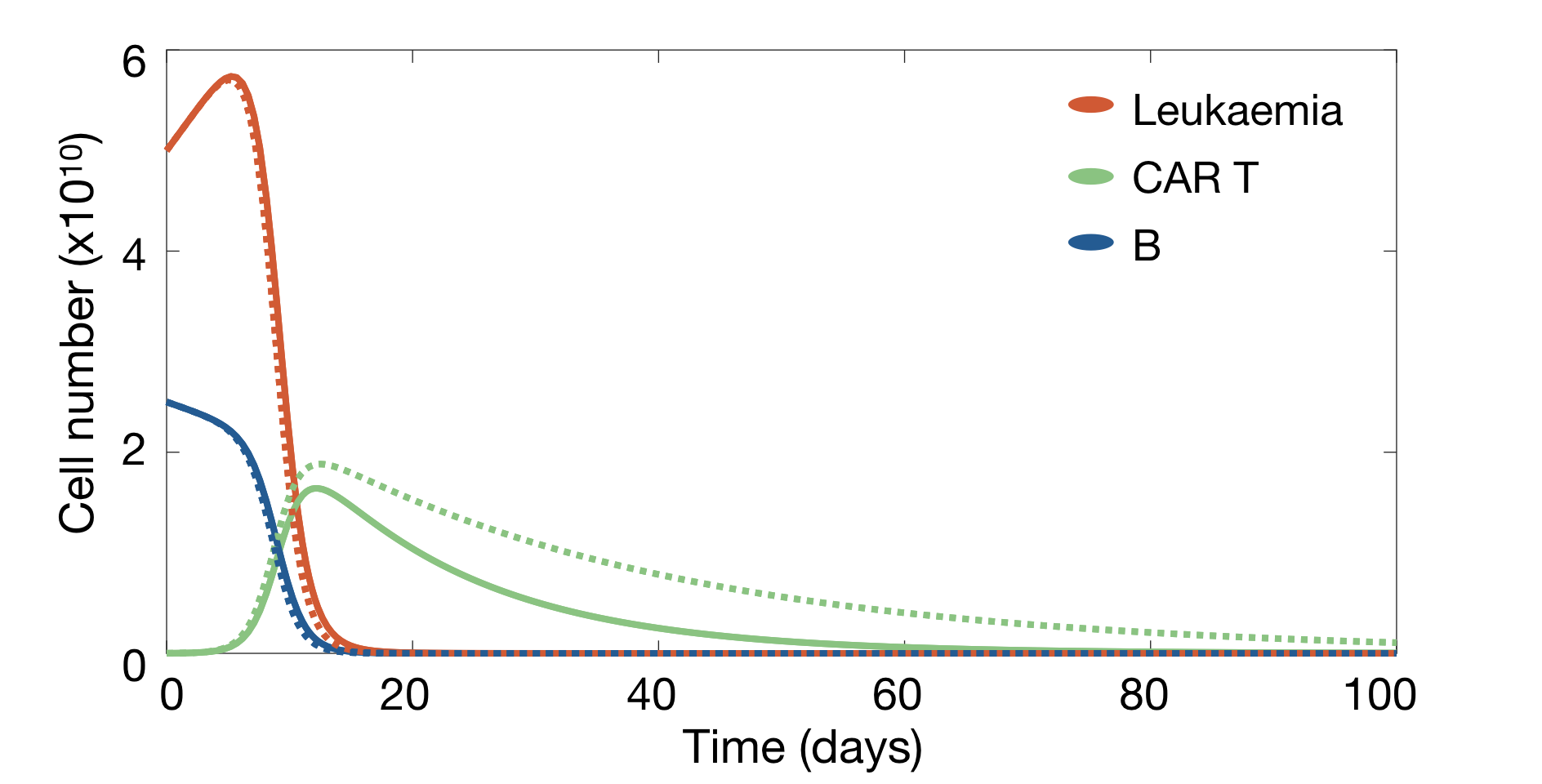}
	\vspace*{-3mm}
	\caption{\textbf{CAR T cell persistence when varying its lifetime $\boldsymbol{\tau_C}$}. Dynamics of the { leukaemic cell} (red), B-cell (blue) and  CAR T cell (green) according to model Eqs.(\ref{model3}). Solid and dashed curves correspond to $\tau_C = 14$ days and $\tau_C = 30$ days, respectively, with the rest of parameters and initial data as in Figure~\ref{prima}.}
\label{modeltau}
\end{figure}

The recent clinical study~{\blue\cite{NatureMedicine2019}} showed much longer persistence of the CAR T cells when their mean lifetime was increased to $\tau_C$ = 30 days, larger than the more common value $\tau_C$ = 14 days. We simulated the dynamics of Eqs. (\ref{model3}) for both values of $\tau_C$. An example is shown in Fig. \ref{modeltau}. While the B-cells and { leukaemic} cells { exhibited similar behaviour}, CAR T cells showed a much longer persistence in line with the clinical observations.
\par

\begin{figure}[t!]
	\centering
	\includegraphics[width=0.9\columnwidth]{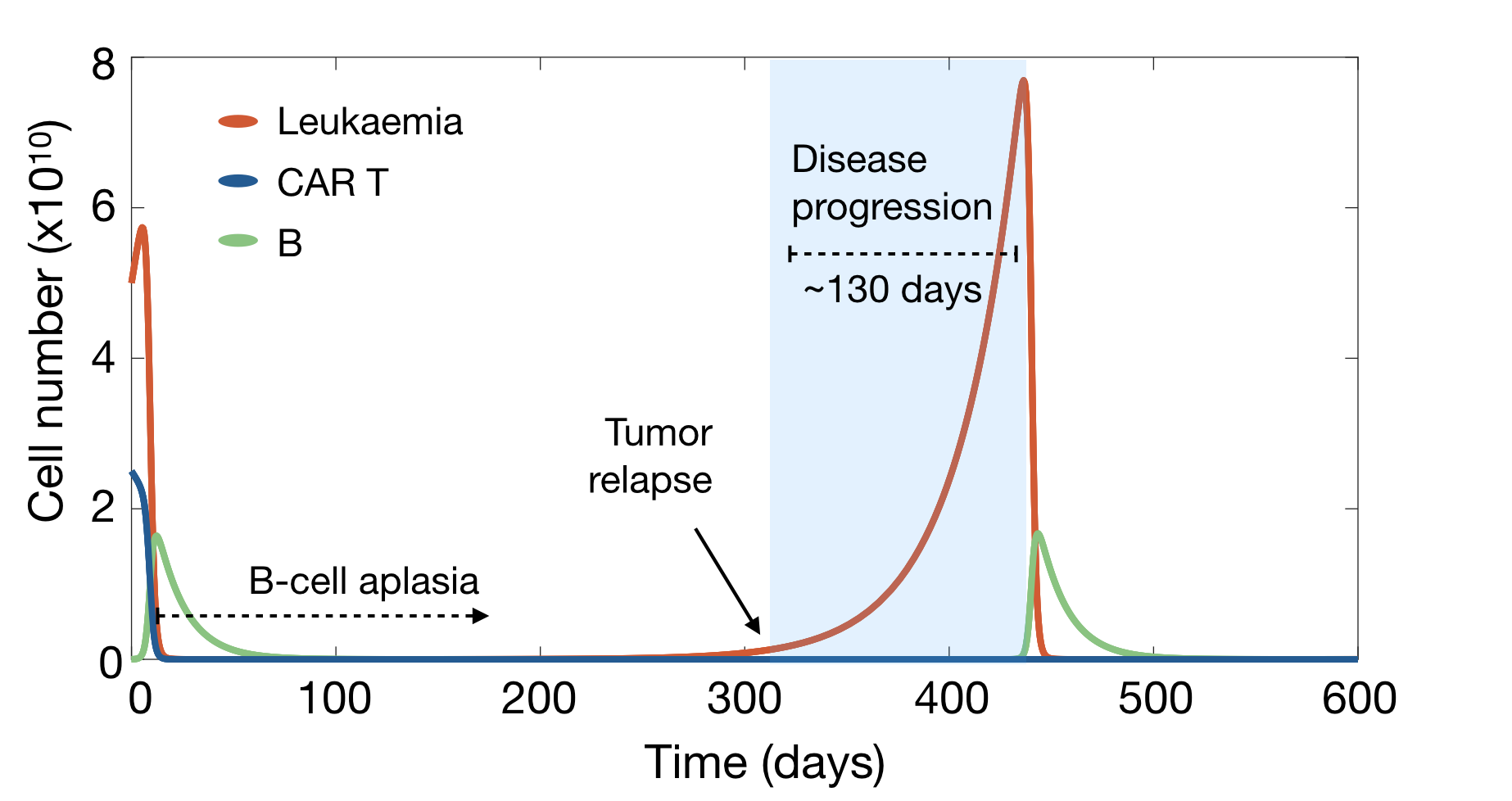}
	\vspace*{-2mm}
	\caption{\textbf{CD19$^+$ relapses could be a dynamical phenomenon}. Long-time dynamics of Eqs.~(\ref{model3}) for { leukaemic} (red), B (blue) and CAR T (green) cells in the time interval [0,600] days, displaying a CD19$^+$ relapse as the result of predator-prey type dynamics {\em in silico}. Parameters are as in Figure~\ref{prima}. The shaded area indicates the time interval in which the disease would be progressing without further interventions. Notice the subsequent emergence of CAR T cells after the CD19$^+$ cell relapse.}
\label{relapse}
\end{figure}

\subsection{CD19$^+$ relapses could be a dynamical phenomenon} 

We performed simulations of Eq. (\ref{model3}) for longer timescales (with parameters as in Fig. \ref{prima}) and observed a long-time relapse (see Fig.~\ref{relapse}) at about one year after infusion, in what would be a CD19$^+$ relapse. { Leukaemic} growth continued for several months but finally there was an outgrowth of CAR T cells after the relapse that was able to control the disease. This is an important nonlinear dynamical phenomenon that could help explain some CD19$^+$ relapses. 
\par

When $B \sim 0$, { as after} CAR T cell expansion, Eqs.~(\ref{model3}) become the well-known Lotka-Volterra predator-prey mathematical model. That model gives rise to periodic oscillations corresponding to ecological cycles that have been observed both in ecosystems~{\blue\cite{libro}} and in experimental models~{\blue\cite{NatureLotka}}. In our present case, the period of the cycles would be related to the { cancer} relapse time. This period { has been previously described as showing} a complex dependence on a conserved quantity, ${ \mathcal{K}}$~{\blue\cite{Period}} { (having in our case units of s$^{-1}$)}, and given by
\par
\begin{multline}
{ \mathcal{K}} = \alpha C - \rho_{ L} \log\left(\frac{\alpha C}{\rho_{ L}} \right)+ \rho_C{ L} -\frac{1}{\tau_C}\log\left( \rho_C \tau_C { L}\right) \sim \rho_C \left({ L}_0 + B_0\right).
\end{multline}
The conserved quantity $ { \mathcal{K}} \gg 1$, can be approximated by the asymptotic formula~{\blue\cite{Asymptotic}}, { which} in our case yields the period $\mathcal{{ L}}$ of oscillations
\begin{eqnarray}
\mathcal{{ L}}({ \mathcal{K}}) &=& \frac{\rho_C \tau_C({ L}_0+B_0)}{\rho_{ L}} +  \frac{1}{\rho_{ L}}\log \left[\rho_C\tau_C\left( { L}_0+B_0\right)\right] \nonumber \\
&+& \tau_C\log\!\left[\frac{\rho_C\left( { L}_0+B_0\right)}{\rho_{ L}}\right] + O\!\left( \frac{\log E}{E} \right) \sim \frac{\rho_C \tau_C \left({ L}_0+B_0\right)}{\rho_{ L}}.
\label{periodT}
\end{eqnarray}
Because of the approximations involved, Eq.(\ref{periodT}) should only be taken as an order-of-magnitude estimate for the { cancer} relapse time. 
However, it is interesting that longer lifetimes of the CAR T cells resulted in longer relapse times according to Eq.~(\ref{periodT}). This could be the reason why so few CD19$^+$ relapses were observed in the recent trial~{\blue\cite{NatureMedicine2019}}, with $\tau_C$ = 30 days, much longer than the more common value $\tau_C \sim$ 14 days.
\par

A very intriguing question is { whether} those relapses could resolve spontaneously due to the predator-prey type competition between the CAR T and { leukaemic} cells. Owing to the long progression { time, this could pass unnoticed, since after progression, other therapeutic actions} would be taken, such as { haematopoietic} transplants, before allowing the CAR T to appear again. 
\par

Although our simulations { point to} a potentially interesting scenario, the model given by Eqs. (\ref{model3}) is { not good for studying long-term} phenomena. One of the missing biological processes in Eqs. (\ref{model3}) is the potential contribution of B-cell production in the bone marrow from CD19$^{-}$ { haematopoietic} stem cells to the maintenance of a pool of CAR T cells. Thus, to get a more realistic insight { into} the dynamics, we simulated Eqs.~(\ref{model2}) { for the same virtual patients
 and biologically reasonable parameters}  $\beta$ = 0.1, $\tau_I$ = 6 days (see Table \ref{table1}) and different values for the production of B cells in the bone marrow embodied by $I_0$. 
\par

\begin{figure}[t!]
	\centering
	\includegraphics[width=0.9\columnwidth]{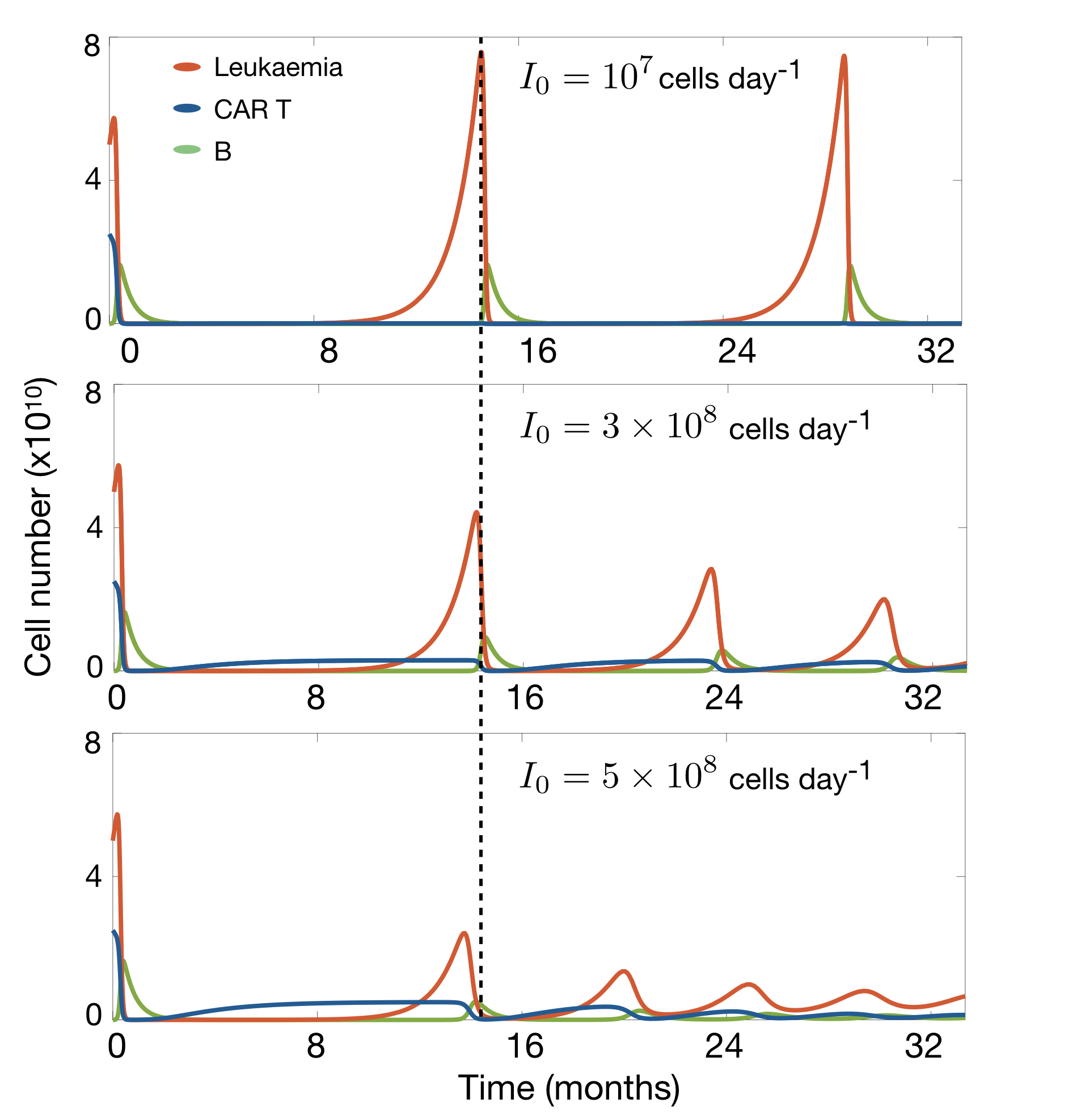}
	\vspace*{-3mm}
	\caption{\textbf{{ Long-time dynamics of virtual patients predicted by} Eqs.~(\ref{model2})}. Parameters are $\alpha = 4.5\times 10^{-11}$  cell$^{-1}$ day$^{-1}$, 
$\beta = 0.1$, $\tau_C = 14$ days, $\tau_B = 60$ days, $\rho_{ L} = 1/30$ day$^{-1}$, $\rho_C = 0.25 \alpha$, 
 $C_{\text{50}}$ = 10$^\text{9}$ cells, $\tau_I= 6$ days. The subplots show the dynamics of the { leukaemic cell} (red), B-cell (blue) and  CAR T cell (green) compartments.
 (a) Case $I_0 = 10^{\text{7}}$ cell day$^{-1}$. (b) Case $I_0 = 3 \times 10^{\text{8}}$ cell day$^{-1}$, (c) Case $I_0 = 5 \times 10^{\text{8}}$  cell day$^{-1}$.}
\label{model2example}
\end{figure}
 
One set of examples is shown in Figure~\ref{model2example}. The more realistic model given by Eqs. (\ref{model2}) still { presents first relapse at a time independent} of the choice of the flux $I_0$. However this parameter influenced the post-relapse dynamics. For values of $I_0$ smaller than approximately  10$^\text{7}$ cells/day, which is the typical number of injected CAR T cells~{\blue\cite{Hartmann2017}}, there were no substantial changes in the dynamics with subsequent relapses following a periodic pattern. Larger values of $I_0$  led to relapses of B cells before the relapse of { leukaemic} cells and later of CAR T cells. Also, the relapse dynamics { were} in line with damped oscillations, with both { leukaemic} and CAR T cell relapses having smaller amplitudes. Interestingly, our model Eqs. (\ref{model2}) predict that a significant increase of B cells could be used as a potential clinical biomarker indicative of subsequent { cancer} relapse.
\par


\subsection{{ CAR T cell} reinjection may allow the severity of relapse to be controlled} 

In the framework of our modelling approach { leukaemia} relapses would be transient. However, { the long potential duration of such relapses} would require further intervention { to prevent patients from suffering undesirable harm during such periods}. An interesting question is { whether it would be possible to control these recurrences} by acting on the { leukaemia} by reinjecting CAR T cells{ , and what the appropriate timings and doses for that intervention would be}.
\par

\begin{figure}[t!]
	\centering
	\includegraphics[width=1\columnwidth]{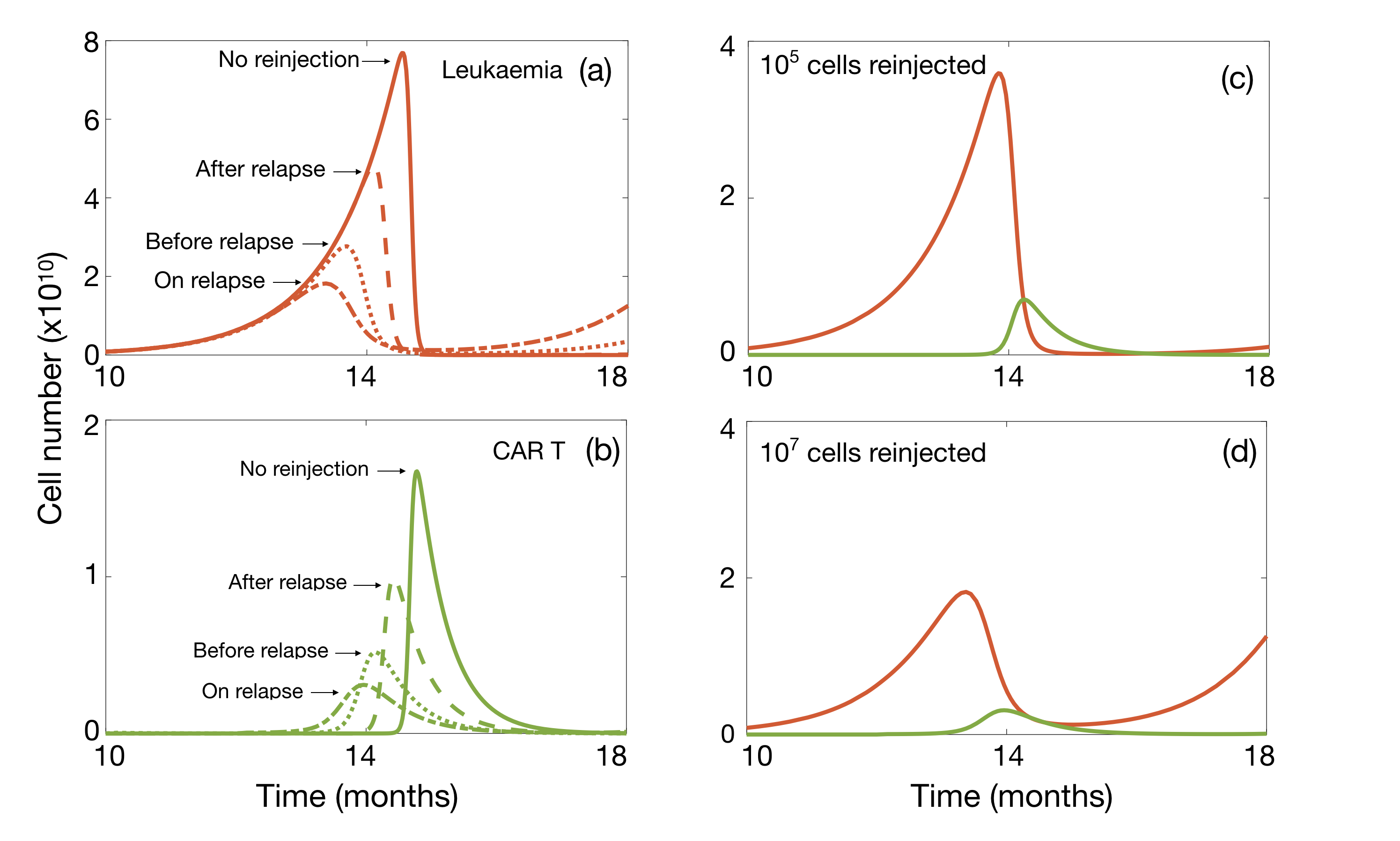}
	\vspace*{-3mm}
	\caption{\textbf{CAR T cell reinjection may allow { severity of relapse to be controlled}}. Simulations of Eqs.(\ref{model3}) for parameter values $\alpha = 4.5\times 10^{-11}$ cell$^{-1}$ day$^{-1}$, $\beta = 0.1$, $\tau_C = 14$ days, $\tau_B = 60$ day, $\rho_{ L} = 1/30$ day$^{-1}$, $\rho_C = 0.25 \alpha$, $C_{\text{50}} = 10^\text{9}$ cells, $\tau_I= 6$ days, and $I_0 = 2\times 10^\text{5}$ cell day$^{-1}$. All subplots show the dynamics of { leukaemic} (red) and CAR T (green) cells upon reinjection of CAR T cells. Cases (a) and (b) display the dynamics of (a) { leukaemic} and (b) CAR T cells, respectively, for doses of $C = 10^\text{7}$ cells administered at times: $t = 300$ days (dash-dot line),  $t = 360$ days (dotted line) and $t = 415$ days (dashed line) in comparison with the dynamics without reinjection (solid line). Subplots (c) and (d) illustrate the combined dynamics of { leukaemic} and CAR T cells after reinjection of: (c) $C = 10^\text{5}$ cells and (d) $C = 10^\text{7}$ cells at  $t =360$ days.}
	\label{reinjection}
\end{figure}

Using the mathematical model~(\ref{model2}), we simulated the reinjection of $C = 10^7$ CAR T cells at different times: before relapse ($t = 300$ days), { at} relapse ($t = 360$ days), and after relapse ($t= 415$ days) and compared the outcome with the case without reinjection. An example is shown in Figure~\ref{reinjection}(a,b). 
\par

Significant reductions of both the peak { leukaemic} cell number and relapse duration were obtained, the best results { being when} reinjection was performed on relapse. Thus, our {\em in silico} results suggest that the early reinjection of CAR T cells in a CD19$^+$ B-leukaemia relapse could { reduce disease load} and help in early control of the disease. This has interesting implications since, after relapse { is detected}, CAR T preparation requires blood extraction, apheresis, T cell modification and expansion {\em ex vivo}, and finally patient infusion. { In clinical} practice this process takes { from} three to six weeks. From the practical point of view, a possibility { for increasing the speed of action} after { leukaemic} cell identification would be to freeze and keep { some of the CAR T cells initially obtained} so that they could be reinjected and aid { in early} control of the disease.
\par

We also studied the effect of the number of CAR T cells injected at the optimal time. An example is shown in Figure~\ref{reinjection}(c,d). The effects of a very small infusion of  $C=10^5$ cells { are compared with those of a} more standard dose of  $C=10^7$ cells. The number of T cells injected { affected} the outcome. This was different from our previous observation that
the number of CAR T cells injected initially did not affect the  treatment outcome. The reason is that, initially, there are many targets, both { leukaemic} and B cells, { allowing for} a huge expansion of the CAR T population. However, on relapse, the target population is smaller and a larger initial number of CAR T cells helps in making the expansion process faster.
\par

\subsection{Model \eqref{model2} predicts { a scenario leading to zero} { leukaemic} cells}

The analysis of the { non-negative} equilibrium points of system \eqref{model2}, { set out} in \ref{AppendixC}, shows the possibility { of reaching} ${ L}=0$ { after starting with a non-zero} { leukaemic} cell population. Using the parameters given in Table \ref{table1}, we observe that there exist ranges for the ratio $k$ and the bone marrow B cell production $I_0$ where one of the equilibrium points, $P_3$ (a focus), is asymptotically stable for different values of $\alpha$. Hence, one of its associated eigenvalues will be real (see Figure~\ref{autovalores} in \ref{AppendixC}) while the other two eigenvalues will be complex conjugate (see Figure \ref{autovalores2} in \ref{AppendixC}). Figure~\ref{model2extinction} illustrates an example with different initial conditions and the same set of parameters for them. It should be pointed out that in all cases shown in Figure~\ref{model2extinction}, both the CAR T and the B cells remain at non-zero levels (of the order of $5\times10^{8}$ and $10^{9}$, respectively) when the { leukaemic} cell population effectively becomes extinct (${ L < 1}$.)
\par

\begin{figure}[t!]
	\centering
	\includegraphics[width=0.7\columnwidth]{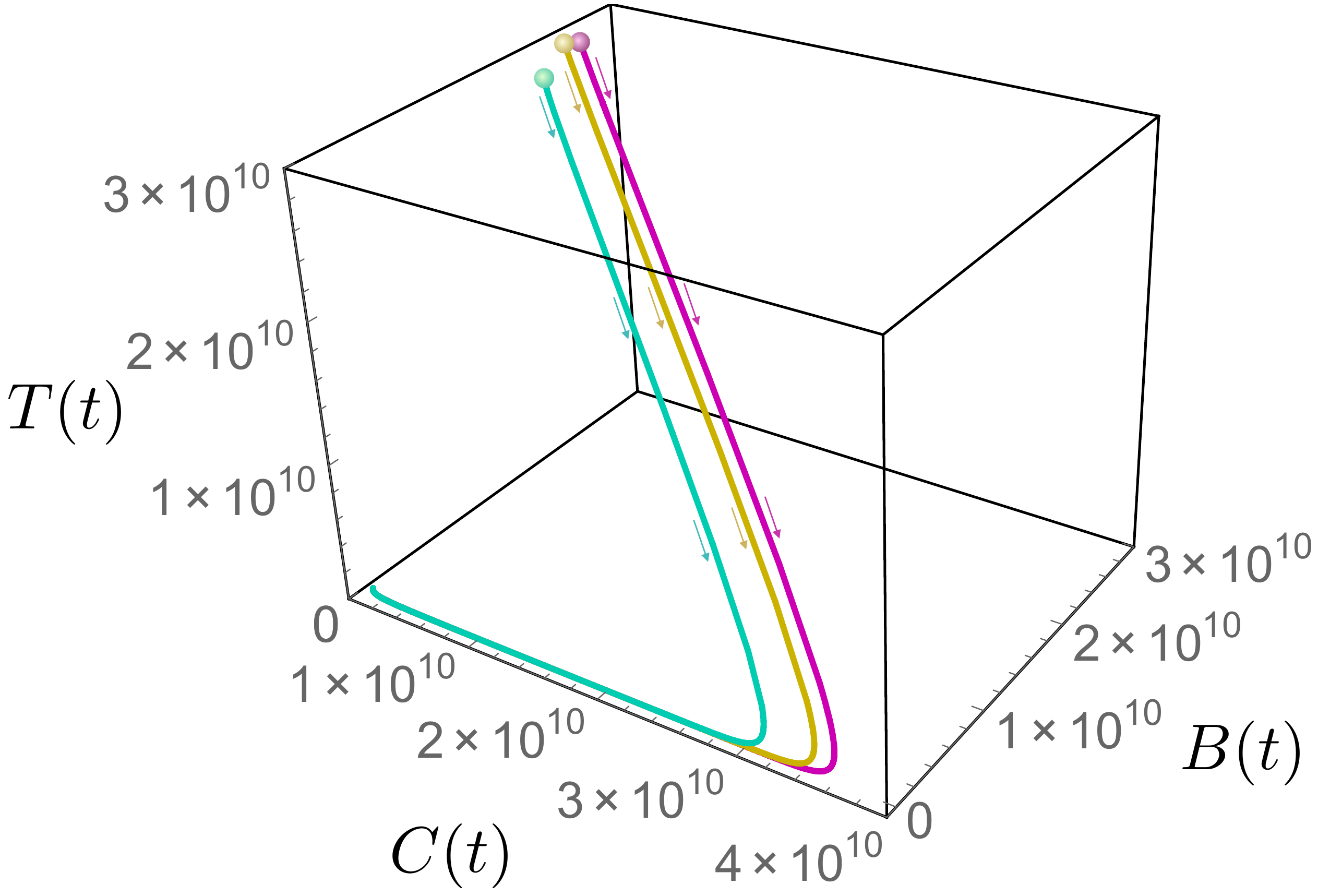}
	\vspace*{-3mm}
	\caption{\textbf{Routes to { leukaemic} cell extinction.} Phase portrait of system \eqref{model2} showing three orbits with different initial conditions ({ coloured} dots) that lead to { leukaemic} extinction. Parameters for all orbits are $\alpha = 5\times 10^{-11}$  cell$^{-1}$ day$^{-1}$, $\beta = 0.1$, $\tau_C = 20$ days, $\tau_B = 40$ day, $\rho_{ L} = 1/45$ day$^{-1}$, $\rho_C = 0.7 \alpha$, 
		$C_{\text{50}}$ = 10$^\text{9}$ cells, $\tau_I= 2.4$ days, and $I_0 = 2 \times 10^{\text{8}}$ cell day$^{-1}$.}
	\label{model2extinction}
\end{figure}

These results are interesting as they suggest that there is a range of biologically relevant parameters where the { cancer} may eventually disappear. Our simulations indicate that the larger $I_0$, $k$ and $\alpha$, the more likely it is that ${ L} = 0$ { can be reached}. { In particular, if $I_0$ is increased}, this would imply higher production of { both} B and CAR T cells, due to the contribution of immature B cells from the bone marrow, resulting in a { greater chance of eradicating} the { leukaemic cells}. This scenario provides another proof of { the} concept that the mathematical model put forward here can be useful in the { clinical setting} and may trigger new exploratory pathways.
\par

\subsection{Sensitivity analysis}

\begin{figure}[t!]
	\hspace*{-3mm}
	\includegraphics[width=1.03\columnwidth]{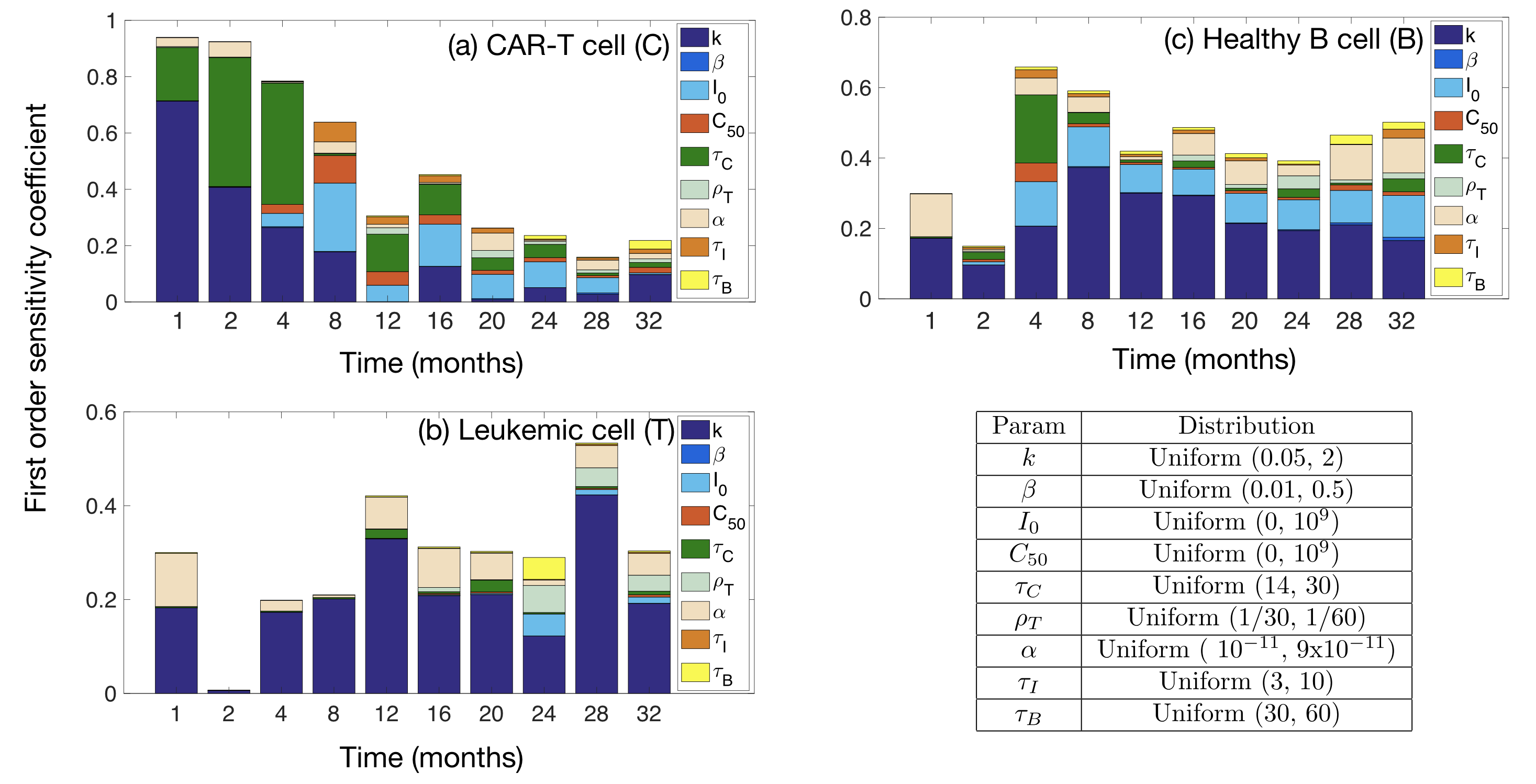}
	\vspace*{-5mm}
	\caption{Sensitivity analysis of system \eqref{model2} to identify the influence of the model parameters on the solutions for (a) CAR T, (b)  { leukaemic}  and (c) B cells. The parameter ranges studied and distributions used are displayed in the lower-right table.}
	\label{sensibilidad}
\end{figure}

A sensitivity analysis was carried out to identify the model parameters { with the greatest influence} on the equilibria for CAR T,  { leukaemic}  and B cells are $I_0$, $\alpha$ and $k$. To do so, we calculated the first-order sensitivity coefficient using Sobol's method~\cite{Saltelli2010} to measure the fractional contribution of a single parameter to the output variance. Using a priori information on the parameters,  we defined the distribution functions in the table shown in Figure~\ref{sensibilidad}. We generated a set of parameters of size 1000 to calculate the sensitivity indices. The results of the sensitivity analysis of  Eqs. \eqref{model2} are { shown} in Figure~\ref{sensibilidad}.

The results show that the parameters { with the greatest influence} on the solutions for CAR T,  { leukaemic}  and B cells are $k$, $\alpha$, $\tau_{C}$ and $I_{0}$. However, their impact varies depending on the specific cell compartment and time. For CAR T cells, $k$ and $\tau_{C}$ are the most { significant} during the first four months after injection. For longer times, $I_{0}$ becomes the { most important}. For  { leukaemic}  cells, $k$ and $\alpha$ are the most influential parameters, both during the first weeks of the CAR T cell treatment and later during relapse. For healthy B cells, $k$ and $\alpha$ are the most relevant parameters during the first weeks, but later on, on relapse, $I_{0}$ also becomes important. Such parameter dependences are also suggestive in order to target specific mechanisms that would allow partial control over them. 

\par

\section{Discussion}

Cancer immunotherapy with CAR T cells is a promising therapeutic option already available for B cell haematological cancers. There has been a growing interest { in} the mathematical descriptions of immunotherapy treatments in cancer, particularly aimed at CAR T cells~{\blue\cite{Kimmel,Rodrigues,Anna}}, but none of them have addressed the specificities of CAR T-based immunotherapies for the case of acute lymphoblastic leukaemias (ALL) on the light of available clinical experience.
\par

In this work, we put forward a mathematical model incorporating the main cell populations involved in the growth of ALL. The model included not only leukaemic clones and CAR T cells, but also the haematopoietic compartment that would be responsible for the persistence of CAR T cells by the continuous generation of CD19$^+$ progenitors from CD19$^-$ stem cells.
\par

One simplified version of the full model already allowed us to describe the clinical evolution of B ALL in the first months after CAR T injection yielding explicit formulas of clinical added value such as the maximum number of CAR T cells that can be reached. Also, it provided a rational support to several clinical observations. Interestingly, the model predicted the possibility of CD19$^+$ relapses being dynamical phenomena resembling predator-prey oscillations. The more complex mathematical models were used to confirm this dynamic and to further give support for therapeutically rechallenging the  { leukaemia}  with CAR T cells in CD19$^+$ relapses. 
\par

{ Although our model, { which} includes essential components driving the CAR T cell and  { leukaemic}  cell dynamics, is shown to provide a description of the response to treatments, there are several limitations in extending our analysis to longer times and to the study of resistances. }First of all, we did not study the case of CD19$^-$ relapses. That study would require a different type of modelling in line with previous mathematical frameworks accounting for the development of other types of resistances under the evolutive pressure of treatments~{\blue\cite{SciRepArturo}}. Also, our description of CD19$^+$ relapses was based on a continuous model, that is not designed to faithfully capture the regime where the numbers of predators (CAR T cells) and { prey} ({ leukaemic}  and B cells) are low. It has been discussed { how} in those scenarios the situation is significantly more complex and may require different types of approaches, such as those based on stochastic birth-death processes or even fully discrete models accounting for different body compartments. However, the fact that the bone marrow compartment would provide a continuous flux of B cells leading to the maintenance of a pool of activated CAR T cells, may keep the system out of the very low densities that could make the continuous model fail.
\par

B-cell lymphomas are usually treated with rituximab to provide a permanent lymphodepletion { also affecting} HSC. Although lymphomas are different diseases, it is interesting to { note} that CAR T cells { would} be expected to persist for shorter times in that scenario due to the lack of continuous stimulation for maintenance of a pool of these cells in the bone marrow. Thus, { it is expected} that treatments would be less effective for these  { cancers} .  
\par

{ It is interesting to point out that in homeostasis there is a population of T cells, the T regulators (Tregs), that control the total number of T cells and have a role in limiting { autoimmune} processes. We did not incorporate Tregs in our mathematical description. This could be a good approximation for the first weeks { of the CAR T cell} expansion because the initial lymphodepletion also affects Tregs. However, after the first 3-4 weeks, this population will be able to expand again and has an effect on a faster reduction of the total CAR T cell load. { There is not much data available} on the dynamics of Treg cells and their reconstitution after the CAR T cell peak. We plan on accounting for this population in future works.}

\par

The results obtained using the reduced mathematical model show that the number of CAR T cells initially injected does not affect the subsequent dynamics. Since at the outset CAR T cells do have a huge {  \em in vivo} target pool, including  { leukaemic}  and healthy B cells allowing them to expand, even small doses of properly functioning immune cells would lead to a response. Thus, according to our modelling approach, it may be better to store (freeze) part of the cells generated so that they could be ready for later  { leukaemia}  rechallenging in case of a CD19$^+$ relapse. There, the combination of a fast action after the detection of the disease and the injection of a substantial number of CAR T cells would be clinically relevant according to our mathematical model-based predictions. The reason is that a prompt action would allow both for a reduced growth of the disease and for a smaller toxicity of the disease, thus reducing risks for the patient such as CRS and ICANS. The rationale behind the injection of larger CAR T loads on relapse is that the target population would be smaller in general than at the start of the treatment. Moreover, our model implies that a periodic treatment with CAR T cells to avoid relapse would be quite ineffective, since they would not be expected to expand well unless there is a substantial target population. 
\par

Our mathematical model allowed us to obtain an estimate for the very relevant parameter of the relapse time, that would be the optimal time to perform the re-injection of CAR T cells. The estimation obtained by Eq. (\ref{periodT}) shows that the relapse time depends on the parameters related to CAR T cells ($\rho_{C}$ and $\tau_{C}$), the growth rate $\rho_{{ L}}$ of  { leukaemic}  cells and their density at the beginning of treatment. In the framework of our continuous model all  { leukaemias}  experience a relapse, however longer relapse times may correspond in some simulation runs to  { leukaemias}  having intermediate densities { that are} unrealistically low for very long times. Thus, we may expect that  { leukaemias} with quite long relapse times, according to our modelling approach, { would} never relapse. 
\par

We may act therapeutically on $\rho_{C}$ and $\tau_{C}$ by designing CAR T cells with higher stimulation ratios and longer persistence. However, it is important to note that the first parameter would be expected to influence the initial treatment toxicity (CRS and ICANS), thus probably, increasing the second would be a { better} option, perhaps by increasing the fraction of CD8$^+$ memory T cells in the CAR T cell pool. Contradictory { evidence has been found in this regard}, although in this case the endpoint to consider would be the number of CD19$^+$  { leukaemia}  relapses.
\par

{ One of the most impactful findings in this research was the very relevant role of the flux of generation of CD19$^+$ progenitors from CD19$^-$ haematopoietic stem cells, $I_0$. Although these stem cells are known to be a very small population (typically around 1\% of all cells in the bone marrow), { what matters most} is the flow { into} the compartment of B-cells. We have recently constructed a model of B-cell lymphopoiesis based on flow-cytometry data that can provide typical populations in equilibrium \cite{Salvi} but the dynamics are difficult to estimate since what is { most} relevant is { what the flow of progenitors would be} when the CD19$^+$ compartment is empty. We are now collecting data on bone marrow reconstitution after haematopoietic stem cell transplants in order to estimate these parameters and have more precise estimates of $I_0$. Our finding { also opens} the interesting possibility of considering the pharmacological stimulation of the 
	process of stem cell asymmetric division and differentiation. This would { lead} to an increase in $I_0$ and thus { raise} the possibility of cancers being completely removed.}

\par
Our mathematical model also allows us to pose another interesting hypothesis. B-ALL is a field where substantial { progress has} been made by designing initial intensive treatment regimes combining different types of cytotoxic chemotherapies. On the basis of our mathematical model, and leaving aside the important { economic} costs, one would expect that substantially less aggressive chemotherapy regimes could be quite effective after CAR T cell injection to eliminate the residual disease (mainly by greatly reducing the $\rho_C$ parameter). { This} strategy would also be beneficial to control CD19$^-$ relapses. In { this} case one should balance the side effects of current protocols versus a combination of CAR T cells with a reduced chemo infusion. The main limitation of this approach would be the sustained B-lymphodepletion provoked by the immunotherapy treatment, but one can envision autologous B stem cell transplants after {\em in vitro} treatment with CAR T cells to select for CD19$^-$ HSCs.
\par

In conclusion we have put forward a mathematical model describing the response of acute lymphoblastic leukaemias to the injection of CAR T cells. Our theoretical framework provided a mechanistic explanation of the observations reported in different clinical trials. Moreover, it also predicted that CD19$^+$  { leukaemia}  relapses could be the result of the competition between  { leukaemic}  and CAR T cells in an analogous fashion to predator-prey dynamics. As a result, the severity of relapses could be controlled by early rechallenging of the  { leukaemia}  with previously stored CAR T cells.

\section*{Acknowledgement}
This work has been partially supported by the Junta de Comunidades de Castilla-La Mancha (grant number SBPLY/17/180501/000154), the James S. Mc. Donnell Foundation (USA) 21st Century Science Initiative in Mathematical and Complex Systems Approaches for Brain Cancer (Collaborative award 220020450), Junta de Andalucía group FQM-201, Fundación Española para la Ciencia y la Tecnología (FECYT, project PR214 from the University of Cádiz) and the Asociación Pablo Ugarte (APU). OLT is supported by a PhD Fellowship from the University of Castilla-La Mancha research plan.
\appendix

\section{Basic properties of the complete mathematical model: system {\bf \eqref{model1}} }
\label{AppendixA}

We state the following proposition:
\begin{Proposition}
	\label{Prop:1}
	For any { non-negative} initial data $(C(0),$ ${ L}(0),$ $B(0),$ $P(0),$ $I(0))$ and all parameters of the initial value problem given by Eqs. (\ref{model11})-(\ref{model15}) being positive, the solutions for $C(t), { L}(t), B(t), P(t),$ and $I(t)$ exist for all $t>0$, are unique and { non-negative}. 
\end{Proposition}
\textbf{Proof.} We first { prove the non-negativity} of the solutions. Let ${\bf F} = {\bf F}({\bf x})$ denote the vector field representing the right-hand-side of Eqs. (\ref{model11})-(\ref{model15}), with function ${\bf x} \equiv \left( C, { L}, B, P, I\right)$. Also, let ${\bf n}_{j}$ denote the outward normal unit vector to plane $x_{j}=0$, with $j=1,2,\ldots,5$. That is, ${\bf n}_{1} = (-1,0,0,0,0)$ and analogously for other ${\bf n}_{j}$. Consider the scalar products of the ODE system $\frac{d{\bf x}}{dt} = {\bf F}({\bf x})$ with each ${\bf n}_{j}$ and assume that the initial data $(C(0),$ ${ L}(0),$ $B(0),$ $P(0),$ $I(0))$ are positive. Then,  $\frac{d{\bf x}}{dt}{\bf \cdot n}_{1} = {\bf F \cdot n}_{1} = 0$, $\frac{d{\bf x}}{dt}{\bf \cdot n}_{2} = {\bf F \cdot n}_{2} = 0$ and $\frac{d{\bf x}}{dt}{\bf \cdot n}_{4} = {\bf F \cdot n}_{4} = 0$ at hyper-surfaces $C=0$, ${ L}=0$ and $P=0$, respectively. 
Then, the hyper-surfaces $C=0$, ${ L}=0$ and $P=0$ are invariant.

Next, $\frac{d{\bf x}}{dt}{\bf \cdot n}_{5} = {\bf F \cdot n}_{5} = -\frac{1}{\tau_P} P\leq 0$ at plane $I=0$. Finally, $\frac{d{\bf x}}{dt}{\bf \cdot n}_{3} = {\bf F \cdot n}_{3} = -\frac{1}{\tau_I} I\leq 0$ at plane $B=0$. Hence, pieces of hyper-surfaces $\{ I=0\}\cap {\bf R}^{5}_{+,0}$ and $\{ B=0\}\cap {\bf R}^{5}_{+,0}$  are semipermeable inward ${\bf R}^{5}_{+,0}$.

As a result, ${\bf R}^{5}_{+,0}$ is a positively invariant domain for Eqs.~(\ref{model11})-(\ref{model15}). Therefore, { non-negativity} of solutions $\left( C, { L}, B, P, I\right)$ follows.
\par
Since all { parameters in} Eqs. (\ref{model11})-(\ref{model15}) are finite and the right-hand-side of the system is a continuous function in $(C, { L}, B, P, I)$ in the domain ${\bf R}^{5}_{+,0}$, existence of solutions of Eqs. (\ref{model11})-(\ref{model15}) follows from the Cauchy-Peano theorem. Moreover, as the partial derivatives of the right-hand side of the system are also continuous and bounded in ${\bf R}^{5}_{+,0}$, uniqueness follows from the Picard-Lindel\"{o}f theorem. This completes the proof. 

\rule{5pt}{5pt}

\par

\section{Basic properties of the reduced mathematical model: system \eqref{model3} }
\label{AppendixB}

\begin{Proposition}\label{teorema1}
	For any non-negative initial data $(C_{0},{ L}_{0},B_{0})$ and all the parameters of the model being positive, the solutions to Eqs. \eqref{model3} exist for $t>0$, are non-negative and unique.
\end{Proposition}
\textbf{Proof.}  The proof mimics the steps of the proof of Proposition \ref{Prop:1} so the repetitive details are omitted.

\rule{5pt}{5pt}

Before delving into the analysis of system~(\ref{model2}), it is convenient to first understand  the dynamics of a simplified version of~(\ref{model2}) given by Eqs.~(\ref{model3}). We begin by calculating the fixed points and { determining} their stability.
These are the points $P_{1}=(0,0,0)$ and $P_2=(\frac{\rho_{ L}}{\alpha}, \frac{1}{\rho_C\tau_C},0)$.
\par

To { analyse} the stability of these points, we calculate the Jacobian matrix of Eqs.~(\ref{model3}):
{ $$J(C,{ L},B)=
	\begin{pmatrix}
	\rho_C({ L}+B)-\frac{1}{\tau_C} & \rho_C C & \rho_C C\\
	-\alpha { L} & \rho_{ L}-\alpha C &  0 \\
	-\alpha B & 0 &  -\alpha C -\frac{1}{\tau_B}
	\end{pmatrix}.
	$$}

\begin{itemize}
	\item Equilibrium point $P_{1}=(0,0,0)$. The Jacobian matrix is
	$$J(P_{1})=
	\left( \begin{array}{ccc}
	-\frac{1}{\tau_C} & 0 & 0 \\
	0 & \rho_{ L} & 0 \\
	0 & 0 & -\frac{1}{\tau_B} 
	\end{array}\right),
	$$
	and the eigenvalues are $\lambda_{1}= 1/\tau_C, \lambda_{2}=\rho_{ L}$ and 
	$\lambda_{3}=-1/\tau_B$. Thus, $P_{1}$ is a saddle point and therefore, an unstable equilibrium point.

	\item Equilibrium point $P_{2}=(\frac{\rho_{ L}}{\alpha},\frac{1}{\rho_C\tau_C},0)$.
	
	If we make the following linear change of coordinates $x=C-\rho_{ L}/\alpha$, $y={ L}-1/\rho_C\tau_C$ and $z=B$ we move the point $P_2$ to the origin and the system \eqref{model3} becomes
	\begin{eqnarray}
	\frac{dx}{dt}&=&\frac{\rho_C\rho_{ L}}{\alpha}(y+z)+\rho_C(y+z)x,\\
	\frac{dy}{dt}&=&-\frac{\alpha}{\rho_C\tau_C}x-\alpha xy,\\
	\frac{dz}{dt}&=& -(\rho_t+\frac{1}{\tau_B})z-\alpha xz.
	\end{eqnarray}
	The Jacobian matrix for this point is:
	$$J(P_{2})=
	\left( \begin{array}{ccc}
	0 & \frac{\rho_C\rho_{ L}}{\alpha} & \frac{\rho_C\rho_{ L}}{\alpha} \\
	-\frac{\alpha}{\rho_C\tau_C}  & 0 & 0  \\
	0 & 0  &-\rho_{ L}-\frac{1}{\tau_B} 
	\end{array} \right).$$
	The eigenvalues of the matrix are $\lambda_{1,2}=\pm\sqrt{\frac{\rho_{ L}}{\tau_C}}i$ and $\lambda_{3} = -\rho_{ L}-\frac{1}{\tau_B}$.
	
	Since $\lambda_{1,2}$ are imaginary eigenvalues, this point is a non-hyperbolic point. This means that for the { linearised} system, $P_2$ is a centre, but it is not possible to conclude its stability for the nonlinear system. 
	
	On the other hand, since $\lambda_3<0$, $P_2$ possesses a local stable manifold corresponding to that eigenvalue. Thus, $P_2$ has a local { centre} manifold (corresponding to the eigenvalues $\lambda_{1,2}$) of dimension $2$ and a local stable manifold (corresponding to the eigenvalue $\lambda_3$).
	
	Since $\lambda_3$ is negative, all the orbits starting near the equilibrium point approach the { centre} manifold. It is straightforward (although somewhat tedious) to verify that the { centre} manifold is given by $z=h(x,y)=0$. So, the qualitative behaviour of the local flow can then be determined from the flow of the following system on the { centre} manifold $z=0$: 
	\begin{eqnarray}\label{sist_LV}
	\frac{dx}{dt}&=&\frac{\rho_C\rho_{ L}}{\alpha}y+\rho_Cxy,\\
	\frac{dy}{dt}&=&-\frac{\alpha}{\rho_C\tau_C}x-\alpha xy.
	\end{eqnarray}
	Making the following change of variables
	$$
	t\to\frac{\alpha}{\rho_C\tau_C}\sqrt{a}t,\quad y\to\sqrt{a}y, 
	$$
	where $a=\rho_C^2\rho_{ L}\tau_C/\alpha^2$, $b=\rho_C^2\tau_C/\alpha$ and $m=\rho_C\tau_C$, we obtain
	\begin{equation}\label{eq_sist_1}
	\begin{split}
	\frac{dx}{dt}&=y+\frac{b}{a}xy,\\
	\frac{dy}{dt}&=-x-\frac{m}{\sqrt{a}}xy.
	\end{split}
	\end{equation}
	If we introduce polar coordinates, defined by
	$$
	x=r\cos\theta,\quad y=-r\sin\theta,
	$$
	system \eqref{eq_sist_1} becomes
	\begin{equation}
	\begin{split}
	\dot{r}&={\cal R}(r,\theta),\\
	\dot{\theta}&=1+\Theta(r,\theta),
	\end{split}
	\end{equation}
	where
	\begin{equation}
	\begin{split}
	{\cal R}(r,\theta)&=-\frac{b}{a}r^2\cos^2\theta\sin\theta-\frac{m}{\sqrt{a}}r^2\sin^2\theta\cos\theta,\\
	\Theta(r,\theta)&=\frac{b}{a}r\cos\theta\sin^2\theta-\frac{m}{\sqrt{a}}r\sin\theta\cos^2\theta.
	\end{split}
	\end{equation}
	Thus, we can derive an equation for $r$ as a function of $\theta$ through the differential equation
	\begin{equation}\label{eq_r_ang}
	\frac{dr}{d\theta}=-\frac{\frac{b}{a}r^2\cos^2\theta\sin\theta+\frac{m}{\sqrt{a}}r^2\sin^2\theta\cos\theta}{1+\frac{b}{a}r\cos\theta\sin^2\theta-\frac{m}{\sqrt{a}}r\sin\theta\cos^2\theta}.
	\end{equation}
	In a { neighbourhood} of $r=0$, 
	\begin{eqnarray*}
		\left(1+\frac{b}{a}r\cos\theta\sin^2\theta-\frac{m}{\sqrt{a}}r\sin\theta\cos^2\theta\right)^{-1}= 1+\frac{m}{\sqrt{a}}r\sin\theta\cos^2\theta\\
		-\frac{b}{a}r\cos\theta\sin^2\theta +O(r^2).
	\end{eqnarray*}
	As a result, for small $r$ we { obtain} the following expression as the Taylor series of Eq. \eqref{eq_r_ang}:
	\begin{eqnarray*}
		\frac{dr}{d\theta}&=&-\left[\frac{b}{a}\cos^2\sin\theta+\frac{m}{\sqrt{a}}\sin^2\theta\cos\theta\right]r^2\\
		&+&\left[\frac{b^2}{a^2}\cos^3\theta\sin^3\theta-\frac{m^2}{a}\sin^3\theta\cos^3\theta+\frac{bm}{a\sqrt{a}}\sin^4\theta\cos^2\theta\right]r^3\\
		&-&\left[\frac{bm}{a\sqrt{a}}\cos^4\theta\sin^2\theta\right]r^3 + O(r^4).
	\end{eqnarray*}
	Now, we { use} the method of averaging to { carry out} a change of variable with the effect of reducing the non-autonomous differential equation to an autonomous one. Let the transformation be
	\begin{equation}
	r=\rho+g_1(\theta)\rho^2+g_2(\theta)\rho^3,
	\end{equation}
	with 
	\begin{eqnarray*}
		g_1'(\theta)&=&-\frac{b}{a}\cos^2\theta-\frac{m}{\sqrt{a}}\sin^2\theta\cos\theta,\\
		g_2'(\theta)&=&\left(\frac{b^2}{a^2}-\frac{m^2}{a}\right)\cos^3\theta\sin^3\theta+\frac{bm}{a\sqrt{a}}\sin^4\theta\cos^2\theta-\frac{bm}{a\sqrt{a}}\cos^4\theta\sin^2\theta,
	\end{eqnarray*}
	and formally arrive at the equation
	\begin{equation}
	\frac{d\rho}{d\theta}=0.
	\end{equation}
	At this point, we are in{ a position} to apply the { Centre} Theorem of Lyapunov (see for instance \cite{Hale}) which ensures that when $dr/d\theta$ can formally be transformed to zero, the equilibrium point is a { centre}. Thus, the origin is a { centre and so} the equilibrium point $P_2$ is also a { centre}.

	\begin{figure}[t!]
		\begin{center}
			\includegraphics[width=0.9\columnwidth]{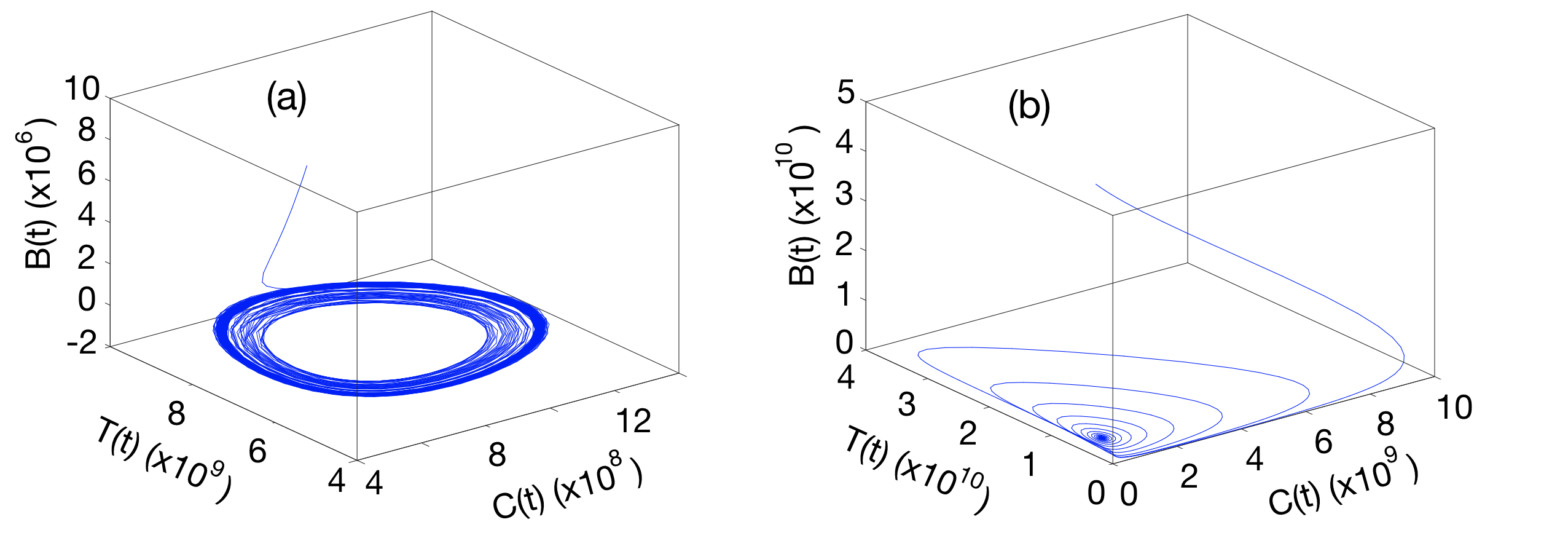}
			\caption{(Left) An orbit of the phase portrait of system \eqref{model3}.  (Right) An orbit of the phase portrait of system \eqref{model2}. The parameters used to calculate such orbits are given in Table 1.
				\label{fig_appendix}}
		\end{center}
	\end{figure}

\end{itemize}

It is also possible to find a Lyapunov function of the system \eqref{eq_sist_1} as
\begin{equation}
V(x,y)=\frac{a}{b}x+\frac{\sqrt{a}}{m}y-\frac{a^2}{b^2}\log\left(1+\frac{b}{a}x\right)-\frac{a}{m^2}\log\left(1+\frac{m}{\sqrt{a}y}\right),
\end{equation}
and $\dot{V}=0$, $\forall (x,y)\in\mathbb{R}^2$. As $\mathbb{R}^{2}_{+}$ is a positively invariant manifold, all the orbits go periodically around $P_2$.
\par
An orbit of the phase portrait for a solution of system \eqref{model3} is shown in Fig. \ref{fig_appendix} (Left).
\par
Finally, let $(C(t), { L}(t))$ be an arbitrary solution of \eqref{model3} for $B=0$, and denote its period by ${\cal { L}}>0$. It is possible to calculate the time average of variables $C$ and ${ L}$ (that is, the number of CAR T and  { leukaemic}  cells, respectively). Dividing the first equation of \eqref{model3} by $C$, the second by ${ L}$ and integrating from $0$ to ${\cal { L}}$ and using the fact that the solutions are periodic, we get
\begin{eqnarray}
0=\log C(t)\big|_{0}^{\cal { L}}=\rho_C\int_{0}^{\cal { L}}{ L}(t)dt-\frac{1}{\tau_c}{\cal { L}},
\end{eqnarray}
and
\begin{eqnarray}
0=\log { L}(t)\big|_{0}^{\cal { L}}=\rho_{{ L}}P-\alpha\int_{0}^{\cal { L}}C(t)dt.
\end{eqnarray}
Hence
\begin{eqnarray}
\frac{1}{{\cal { L}}}\int_{0}^{\cal { L}}C(t)dt&=&\frac{\rho_{ L}}{\alpha},\\
\frac{1}{{\cal { L}}}\int_{0}^{\cal { L}}{ L}(t)dt&=&\frac{1}{\rho_C\tau_C},
\end{eqnarray}
whose values are equal to the equilibrium point $P_2$ for the two first coordinates.

\section{Basic properties of the reduced mathematical model: system \eqref{model2}}
\label{AppendixC}

We have the following proposition for system \eqref{model2}, which is similar to the previous propositions for systems \eqref{model1} and \eqref{model3}, and therefore details of the proof are omitted:
\begin{Proposition}
	\label{Prop:2}
	For any non-negative initial data $(C(0),$ ${ L}(0),$ $B(0))$ and all parameters of the initial value problem given by Eqs. (\ref{model21})-(\ref{model23}) being positive, the solutions for $C(t), { L}(t)$ and $B(t)$ exist for all $t>0$, are unique and non-negative. 
\end{Proposition}

%

%

\vspace{0.3cm}
\noindent
On the other hand, the non-negative equilibrium points of system \eqref{model2} are:  
\begin{itemize}
	\item Equilibrium point $P_1=(C_1^*, { L}_1^*,B_1^*)=\left(0,0,\frac{\tau_B}{\tau_I}I_0\right)$.
	\item Equilibrium point $P_2$ is given by
	\begin{eqnarray*}
		P_2&=&(C_2^*, { L}_2^*,B_2^*)\\
		&=&\left( \frac{\rho_{ L}}{\alpha},\frac{1}{\rho_{C}\tau_C} - \frac{I_{0}\!\left( \tau_{B} +\beta\left( 1+\rho_{ L}\tau_B\right)\!\tau_{I}\right)}{\left( 1+\rho_{ L}\tau_B\right)\!\left( 1+\frac{\rho_{ L}}{\alpha C_{50}}\right)\!\tau_{I}}, \frac{I_{0}\tau_{B}}{\left( 1+\rho_{ L}\tau_B\right)\left( 1+\frac{\rho_{ L}}{\alpha C_{50}}\right)\!\tau_{I}} \right)
	\end{eqnarray*}
	where we assume that
	\begin{eqnarray} \label{Equi2T}
	\frac{1}{\rho_{C}\tau_C} > \frac{I_{0}\!\left( \tau_{B} +\beta\left( 1+\rho_{ L}\tau_B\right)\!\tau_{I}\right)}{\left( 1+\rho_{ L}\tau_B\right)\!\left( 1+\frac{\rho_{ L}}{\alpha C_{50}}\right)\!\tau_{I}}\, .
	\end{eqnarray}
	\item $P_3=(C_3^*, { L}_3^*,B_3^*)=\left(C_3^*,0,\frac{1}{\rho_{C}\tau_{C}} - \frac{\beta I_{0}}{1+C_3^*/C_{50}}\right)$
	where $C_3^*$ is given by
	\begin{eqnarray*} 
		C_3^*&=&-\frac{C_{50}}{2}\left( 1+ \frac{1}{\alpha\tau_{B}C_{50}} - \rho_{C}\beta I_{0}\tau_{C}\right)\nonumber\\
		&+& \frac{C_{50}}{2}\sqrt{\left( 1+ \frac{1}{\alpha\tau_{B}C_{50}} - \rho_{C}\beta I_{0}\tau_{C}\right)^2 + \frac{4\rho_{C}I_{0}\tau_{C}}{\alpha C_{50}}\left(\frac{\beta}{\tau_{B}} +\frac{1}{\tau_I} - \frac{1}{\rho_{C}I_{0}\tau_{C}\tau_{B}}\right)}
	\end{eqnarray*}
	with the following conditions holding
	\begin{eqnarray} 
	\frac{\beta}{\tau_{B}} +\frac{1}{\tau_I} > \frac{1}{\rho_{C}I_{0}\tau_{C}\tau_{B}}, \quad \frac{1}{\rho_{C}\tau_{C}} > \frac{\beta I_{0}}{1+C_{3}^{*}/C_{50}}\, .
	\label{Equi3Conditions}
	\end{eqnarray}
	\item $P_4=(C_4^*, { L}_4^*,B_4^*)=\left(C_4^*,0,\frac{1}{\rho_{C}\tau_{C}} - \frac{\beta I_{0}}{1+C_4^*/C_{50}}\right)$
	where $C_4^*$ is given by
	\begin{eqnarray*} 
		C_4^*&=&-\frac{C_{50}}{2}\left( 1+ \frac{1}{\alpha\tau_{B}C_{50}} - \rho_{C}\beta I_{0}\tau_{C}\right)\nonumber\\
		&-& \frac{C_{50}}{2}\sqrt{\left( 1+ \frac{1}{\alpha\tau_{B}C_{50}} - \rho_{C}\beta I_{0}\tau_{C}\right)^2 + \frac{4\rho_{C}I_{0}\tau_{C}}{\alpha C_{50}}\left(\frac{\beta}{\tau_{B}} +\frac{1}{\tau_I} - \frac{1}{\rho_{C}I_{0}\tau_{C}\tau_{B}}\right)}
	\end{eqnarray*}
	with the following conditions { being satisfied}
	\begin{eqnarray} 
	1+ \frac{1}{\alpha\tau_{B}C_{50}} < \rho_{C}\beta I_{0}\tau_{C}, \  \frac{\beta}{\tau_{B}} +\frac{1}{\tau_I} < \frac{1}{\rho_{C}I_{0}\tau_{C}\tau_{B}} , \ \frac{1}{\rho_{C}\tau_{C}} > \frac{\beta I_{0}}{1+C_{4}^{*}/C_{50}}\, . \qquad
	\label{Equi4Conditions}
	\end{eqnarray}
\end{itemize}

Regarding the study of the stability of the equilibrium points, we obtain the following conclusions:
\begin{itemize}
	\item The eigenvalues of $P_1$ are 
	$$
	\lambda_1=\rho_{ L},\quad \lambda_2=-\frac{1}{\tau_B}, \quad \lambda_3=\frac{I_0\rho_C\tau_B\tau_C-\tau_I+I_0\beta\rho_C\tau_C\tau_I}{\tau_C\tau_I}
	$$
	and therefore $P_1$ is an unstable point (saddle point).
	\item Using the Routh-Hurwitz criterion, it follows that $P_2$ is asymptotically stable for any positive $I_0$ value, assuming that condition \eqref{Equi2T} is satisfied. Since for $I_0=0$ we obtain the system \eqref{model3}, it means that a small perturbation of $I_0=0$, i.e. $I_0=\varepsilon$, with $\varepsilon$ sufficiently small, will transform the centres obtained in the system \eqref{model3} into asymptotically stable foci. Therefore, $I_0=0$  is a bifurcation point since for this value the type of stability changes, and we obtain a Hopf bifurcation. An orbit of the phase portrait for a solution of system \eqref{model2}, for this case, is shown in Fig. \ref{fig_appendix}(right).
	\par
	
	
	\item  Carrying out a stability study, in a general way, for both $P_3$ and $P_4$ is very complex, so we have performed a study within the confines of the parameter range collected in Table \ref{table1}, which are biologically relevant. 
	\par
	We have observed that for all parameters in Table \ref{table1}, $P_4$ has at least one negative component, and thus we do not consider such biologically unfeasible scenarios.
	
\end{itemize}

\begin{figure}[t!]
	\centering
	\includegraphics[width=1.01\columnwidth]{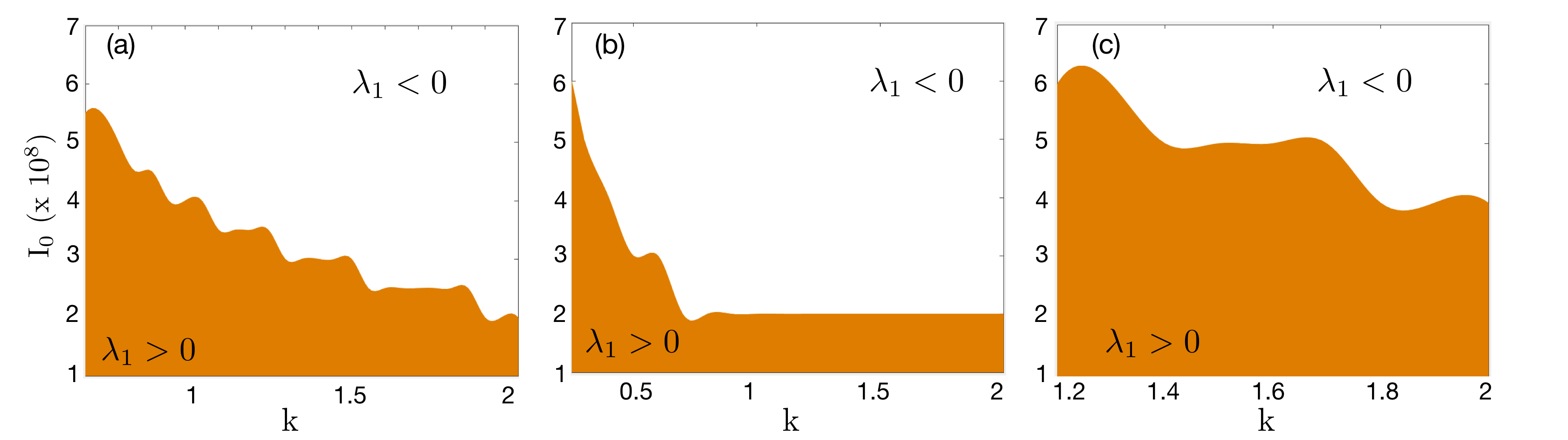}
	\vspace*{-5mm}
	\caption{Values of the sign of eigenvalue $\lambda_1$ for different values of $I_0$ and $k$ for (a) $\alpha=4.5\cdot 10^{-11}$ (b) $\alpha= 10^{-10}$ (c) $\alpha=3\cdot 10^{-11}$. The orange shaded region corresponds to eigenvalue $\lambda_1$ positive.}
	\label{autovalores}
\end{figure}

\begin{figure}[t!]
	\centering
	\includegraphics[width=1\columnwidth]{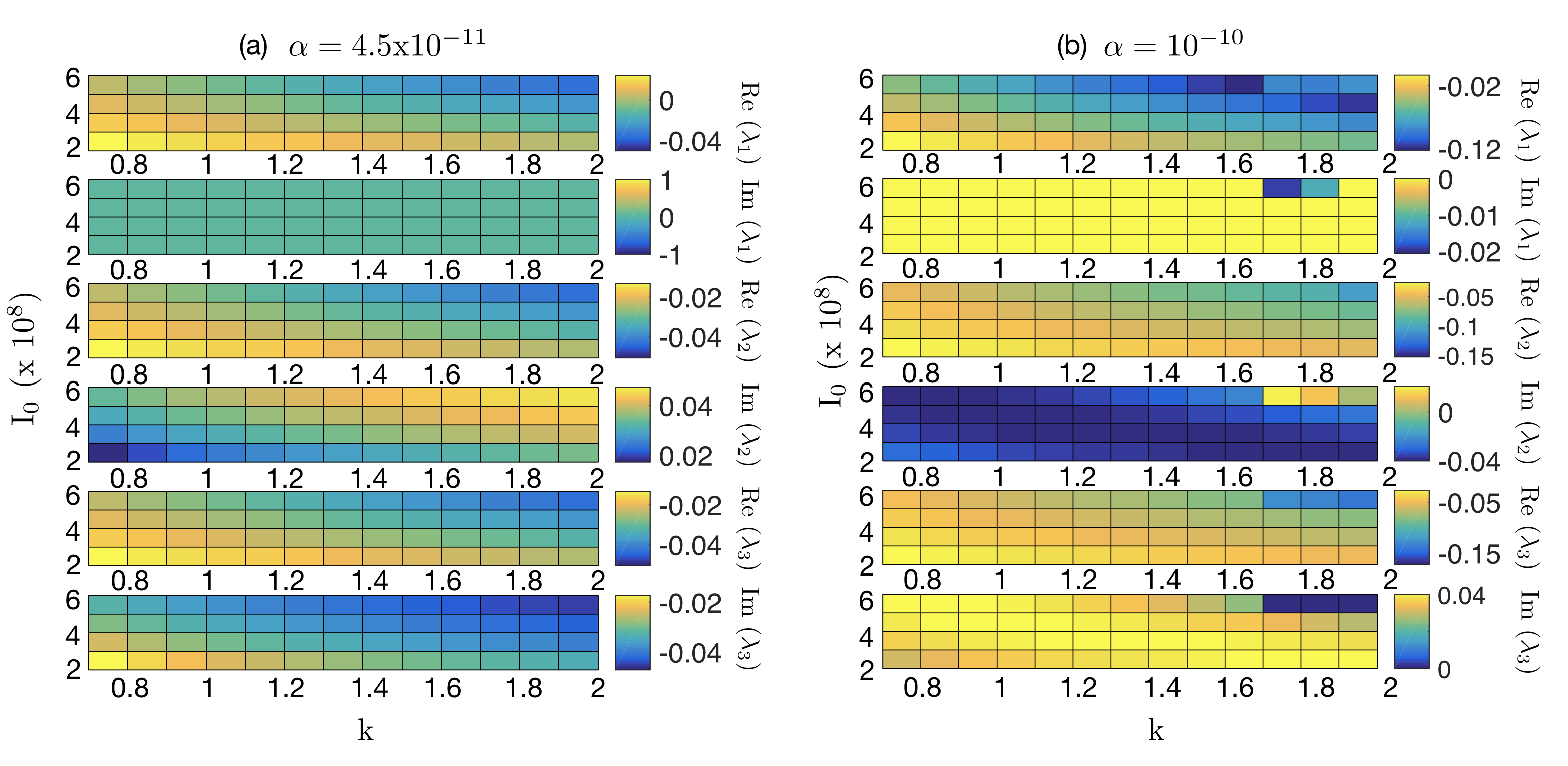}
	\vspace*{-7mm}
	\caption{{ Pseudocolour} plots of the real and imaginary parts of $P_3$ for different values of $I_0$ and $k$ for (a) $\alpha=4.5\cdot 10^{-11}$ and (b) $\alpha= 10^{-10}$}
	\label{autovalores2}
\end{figure}

There exist parameters for $P_3$ for which the components $C_3^*, { L}_3^*$, and $B_3^*$ are all positive and correspond to a point that is an asymptotically stable focus. Figure~\ref{autovalores} shows the region where the real eigenvalue, say $\lambda_1$, is negative or positive, as a function of $I_0$ and $k$ and for different values of $\alpha$. Figure \ref{autovalores2} depicts all the eigenvalues of $P_3$ for different values of $I_0$, $k$ and $\alpha$. { As can} be seen, the real part of the complex eigenvalues $\lambda_2$ and $\lambda_3$, is always negative. On the other hand, the real eigenvalue $\lambda_1$ changes its sign for different values of $I_0$, $k$ and $\alpha$. Then, the stability and instability of $P_3$ is given by the sign of $\lambda_1$.

\section{Analytical formulae for system \eqref{model3}}
\label{AppendixD}

\begin{Lemma}
	\label{Lemma1}
	Let ${ L}_{0}$ and $B_{0}$ denote the initial conditions for the  { leukaemic}  and B cells, which are assumed to be positive numbers. The exact positive solutions of model Eqs.~(\ref{model3}) for the  { leukaemic}  and B cells, denoted by ${ L}={ L}(t)$ and $B=B(t)$, respectively, satisfy for all $t>0$
	\begin{eqnarray} 
	\frac{{ L}(t)}{B(t)} = \frac{{ L}_{0}}{B_{0}}\exp\!\left[ \left( \rho_{ L} + \frac{1}{\tau_{B}}\right)\!t\right]\! .
	\label{model3TB}
	\end{eqnarray}
\end{Lemma}
\textbf{Proof.} The positive solutions to Eqs.~(\ref{model32}) and (\ref{model33}) are given, respectively, by 
\begin{subequations} \label{model3TBproof}
	\begin{eqnarray} 
	{ L}(t) &=& { L}_{0}\exp\!\left( \rho_{ L} t - \alpha\int_{0}^{t} C(s)ds\right)\! ,  \label{model3TBproofT}\\
	B(t) &=& B_{0}\exp\!\left( -\frac{t}{\tau_{B}} - \alpha\int_{0}^{t} C(s)ds\right)\! .  \label{model3TBproofB}
	\end{eqnarray}
\end{subequations}
Formula (\ref{model3TB}) easily follows from Eqs.~(\ref{model3TBproofT}) and (\ref{model3TBproofB}) by calculating their quotient. 

\rule{5pt}{5pt}
\par

\begin{Proposition}
	\label{Prop:4}
	Let $t_\textrm{max}$ denote the time { at which} a local positive maximum of the CAR T cell solution $C=C(t)$ to Eq.~(\ref{model31}) occurs. Then, the positive solutions to Eqs.~(\ref{model32}) and (\ref{model33}) at $t=t_\textrm{max}$ satisfy
	\begin{subequations} \label{model3TBmaxBIS}
		\begin{eqnarray} 
		{ L}(t_\textrm{max}) &=& \frac{{ L}_{0}\, e^{\left( \rho_{ L} + \frac{1}{\tau_{B}}\right)t_\textrm{max}}}{\rho_{C}\tau_{C}\!\left(B_{0} + { L}_{0}\, e^{\left( \rho_{ L} + \frac{1}{\tau_{B}}\right)t_\textrm{max}}\right)}\, ,
		\label{model3TmaxBIS} \\
		B(t_\textrm{max}) &=& \frac{B_{0}}{\rho_{C}\tau_{C}\!\left(B_{0} + { L}_{0}\, e^{\left( \rho_{ L} + \frac{1}{\tau_{B}}\right)t_\textrm{max}}\right)}\, ,
		\label{model3TmaxBIS} 
		\end{eqnarray}
	\end{subequations}
	where ${ L}_{0}$ and $B_{0}$ are the initial conditions for the  { leukaemic}  and B cells, assumed to be positive numbers.
\end{Proposition}
\textbf{Proof.} If $C=C(t)$ has a local positive maximum at $t=t_\textrm{max}$, then $\frac{dC}{dt}=0$ at $t=t_\textrm{max}$. Using Eq.~(\ref{model31}), we get $\rho_C \left( { L}(t_\textrm{max}) + B(t_\textrm{max})\right)  - \frac{1}{\tau_C} = 0$. Thus, ${ L}(t_\textrm{max}) + B(t_\textrm{max}) = \frac{1}{\rho_{C}\tau_{C}}$. Combining this expression with the above formula~(\ref{model3TB}) evaluated at $t=t_\textrm{max}$, Eqs.~(\ref{model3TBmaxBIS}) follow. 

\rule{5pt}{5pt}
\par

\begin{Proposition}
	\label{Prop:5} Let $t_\textrm{max}$ denote the time { at which} a local positive maximum of the CAR T cell solution $C=C(t)$ to Eq.~(\ref{model31}) occurs. Then $t_\textrm{max}$ can be calculated from the implicit relation
	\begin{eqnarray} 
	\log\!\left[ \rho_{C}\tau_{C}\!\left( { L}_{0}\, e^{\rho_{ L} t_\textrm{max}} +  B_{0}\, e^{-\frac{t_\textrm{max}}{\tau_{B}}} \right)\right] - \alpha\int_{0}^{t_\textrm{max}}C(t)dt = 0
	\label{model3tmaxBIS} 
	\end{eqnarray}
\end{Proposition}
\textbf{Proof.} Combining (\ref{model3TBproofT}) and (\ref{model3TBproofB}), and setting $t=t_\textrm{max}$, we get 
\begin{eqnarray} 
{ L}(t_\textrm{max}) + B(t_\textrm{max}) = \left( { L}_{0}\, e^{\rho_{ L} t_\textrm{max}} + B_{0}\, e^{-\frac{t_\textrm{max}}{\tau_{B}}}\right)e^{- \alpha\int_{0}^{t_\textrm{max}} C(s)ds}\! . \label{model3TBproof4}
\end{eqnarray}
Using the fact that ${ L}(t_\textrm{max}) + B(t_\textrm{max}) = \frac{1}{\rho_{C}\tau_{C}}$, Eq. (\ref{model3TBproof4}) can be finally written as (\ref{model3tmaxBIS}).

\rule{5pt}{5pt}
\par

\begin{Proposition}
	\label{Prop:6} 
	Let $C_\textrm{max}$ be the value of the local positive maximum of the CAR T cell solution $C=C(t)$ to Eq.~(\ref{model31}), occurring at time $t_\textrm{max}$. Then,  
	\begin{eqnarray} 
	C_\textrm{max} &=& C_{0} +\frac{\rho_{C}}{\alpha}\!\left( { L}_{0} + B_{0} - \frac{1}{\rho_{C}\tau_{C}}\right) - \frac{1}{\tau_{C}} \int_{0}^{t_\textrm{max}}C(s)ds\nonumber\\
	&+& \frac{\rho_{{ L}}\rho_{C}}{\alpha}\int_{0}^{t_\textrm{max}}{ L}(s)ds - \frac{\rho_{C}}{\alpha\tau_{B}}\int_{0}^{t_\textrm{max}}B(s)ds\, .
	\label{model3CmaxExact}
	\end{eqnarray}
\end{Proposition}
\textbf{Proof.} We first combine Eqs.~(\ref{model3}) in the form
\begin{eqnarray} \label{model3Approx}
\frac{1}{\rho_{C}}\frac{dC}{dt} + \frac{1}{\alpha}\left( \frac{d{ L}}{dt} +  \frac{dB}{dt}\right) = -\frac{1}{\rho_{C}\tau_{C}}C + \frac{\rho_{{ L}}}{\alpha} { L} - \frac{1}{\alpha\tau_{B}}B.
\label{model3Approx}
\end{eqnarray}
Upon integration, we get 
\begin{eqnarray} 
C(t) &=& C_{0} -\frac{\rho_{C}}{\alpha}\!\left( { L}(t) + B(t) - { L}_{0} - B_{0}\right) \nonumber\\
&-&\frac{1}{\tau_{C}} \int_{0}^{t}C(s)ds + \frac{\rho_{{ L}}\rho_{C}}{\alpha}\int_{0}^{t}{ L}(s)ds - \frac{\rho_{C}}{\alpha\tau_{B}}\int_{0}^{t}B(s)ds\, .
\label{model3C(t)Exact}
\end{eqnarray}
Setting $t=t_\textrm{max}$ and { using} ${ L}(t_\textrm{max}) + B(t_\textrm{max}) = \frac{1}{\rho_{C}\tau_{C}}$ in (\ref{model3C(t)Exact}), the result follows.
\rule{5pt}{5pt}
\par

Numerical evaluation of the three integrals { on the right-hand side} of (\ref{model3C(t)Exact}) reveals that, for the parameters { shown} in Table \ref{table1}, { each is smaller} (by at least one order of magnitude) than the second term (note that there is also partial cancellation among the three integrals). Hence, we may approximate (\ref{model3C(t)Exact}) by (\ref{model3CmaxApprox}).

\clearpage

\section*{References}


\end{document}